\documentclass[12pt,aps,floats,showpacs,amssymb,tightenlines]{revtex4}

\usepackage{color}
\usepackage{graphicx}
\usepackage{amsmath}
\usepackage{amsfonts}
\usepackage{amscd}
\usepackage{epsfig}
\usepackage{amssymb}
\usepackage{tabularx}

\begin{document}

\title{Linear stability of the Linet - Tian solution with negative cosmological constant.}
\author{Reinaldo J. Gleiser} \email{gleiser@fis.uncor.edu}

\affiliation{Instituto de F\'{\i}sica Enrique Gaviola and FAMAF,
Universidad Nacional de C\'ordoba, Ciudad Universitaria, (5000)
C\'ordoba, Argentina}

\begin{abstract}

In this paper we analyze the linear stability of the Linet - Tian
solution with negative cosmological constant. In the limit of
vanishing cosmological constant the Linet - Tian metric reduces to a
form of the Levi - Civita metric, and, therefore, it can be
considered as a generalization of the former to include a
cosmological constant. The gravitational instability of the Levi -
Civita metric was recently established, and the purpose of this
paper is to investigate what changes result from the introduction of
a cosmological constant. A fundamental difference brought about by a
(negative) cosmological constant is in the structure at infinity.
This introduces an added problem in attempting to define an
evolution for the perturbations because the constant time
hypersurfaces are not Cauchy surfaces. In this paper we show that
under a large set of boundary conditions that lead to a unique
evolution of the perturbations, we always find unstable  modes, that
would generically be present in the evolution of arbitrary initial
data, leading to the conclusion that the Linet - Tian space times
with negative cosmological constant are linearly unstable under
gravitational perturbations.
\end{abstract}


\maketitle
\section{Introduction}

The Linet - Tian metric \cite{linet}, \cite{tian} is a static
cylindrically symmetric solution of Einstein's equations that can be
interpreted as describing a line source in a vacuum space time with
non vanishing cosmological constant $\Lambda$, which can be chosen
to be positive or negative. In the case of vanishing $\Lambda$ the
metric reduces to a form of the Levi - Civita metric \cite{levi}.

There has been a continued interest in the Linet - Tian metric and in the
construction of models in which the metric is involved. As a few
examples we may cite \cite{bonnor} where it was found that, for a
negative cosmological constant, in the limit where the source
vanishes, one obtains a static, but cylindrically symmetric
anti-de-Sitter universe. General properties were analyzed, for
instance in \cite{daSilva}. The properties of the solutions with
positive cosmological constant, and their extension to higher
dimensions were considered in \cite{griffiths}. The geodesics
kinematics and dynamics in the Linet - Tian metric with negative
cosmological constant were analyzed in \cite{brito}.

On the other hand, the question of the evolution of external fields
or of the stability of gravitational perturbations on these space
times has not received the same attention.  The stability of the
related Levi - Civita metric under linear perturbations that break
the cylindrical symmetry was analyzed in a recent study
\cite{glei1}, finding that the spectrum of allowed frequencies
contains one unstable (imaginary frequency) mode for every possible
choice of the background metric. The main purpose of the present
paper is to analyze to what extent these results are modified by the
presence of a cosmological constant, and therefore, extend the
results obtained in \cite{glei1} to the Linet - Tian space times.

In the case of the Linet - Tian metric, as compared with the Levi -
Civita metric, the first thing to notice is that the structures of
the space times that result for the two possible signs of $\Lambda$
are very different, specially as regards their properties at large
distances from the symmetry axis. In the positive cosmological case,
there is a null horizon at some finite distance from the symmetry
axis, while for negative cosmological constants there is a time like
horizon at an infinite proper distance from the axis. This implies
that in the analysis of the evolution of either external fields, or
small perturbations, one has to consider very different boundary
conditions. In this paper we study the linear stability of the Linet
- Tian solution with negative cosmological constant. The positive
cosmological case will be considered in a separate paper.

In the case of a negative cosmological
constant the Linet - Tian metric may be written in the form,
\begin{equation}\label{LTeq1}
ds^2 = Q^{2/3} \left(-P^{p_1}dt^2+P^{p_2}
dz^2+P^{p_3}d\phi^2\right)+ d\rho^2
\end{equation}
where:
\begin{eqnarray}\label{LTeq2}
   Q(\rho) & = & \frac{1}{\sqrt{3 \Lambda}}\sinh\left(\sqrt{3\Lambda}\rho\right) \nonumber \\
   P(\rho) & = & \frac{2}{\sqrt{3 \Lambda}}\tanh\left(\frac{\sqrt{3\Lambda}}{2}\rho\right)
\end{eqnarray}
and the parameters $p_i$ satisfy,
\begin{eqnarray}\label{LTeq4}
   p_1+p_2+p_3 & = &  0  \nonumber \\
  p_1{}^2 +p_2{}^2 +p_3{}^2 & = &  \frac{8}{3}
\end{eqnarray}
The ranges of the coordinates are $-\infty < t < \infty$, $0 \leq
\rho < \infty$,  $-\infty < z < \infty$, and $0 \leq \phi \leq
2\pi$. The $p_i$ may be parameterized as,
\begin{eqnarray}\label{LTeq3}
   p_1 & = & -\frac{2(1-8 \sigma +4 \sigma^2)}{3-6\sigma+12\sigma^2}  \nonumber \\
  p_2 & = & -\frac{2(1+4 \sigma +8 \sigma^2)}{3-6\sigma+12\sigma^2}    \\
  p_3 & = & -\frac{4(1-2\sigma -2 \sigma^2)}{3-6\sigma+12\sigma^2}  \nonumber
\end{eqnarray}
with $\sigma$ restricted to $0 \leq \sigma \leq 1/2$. In this paper we will
use a parametrization due to Thorne \cite{thorne}, where
$\sigma=\kappa/(2+2\kappa)$, and,
\begin{eqnarray}\label{LTeq5}
   p_1 & = & -\frac{2(1- 2\kappa -2 \kappa^2)}{3(1+\kappa+\kappa^2)}  \nonumber \\
  p_2 & = & -\frac{2(1+4 \kappa +\kappa^2)}{3(1+\kappa+\kappa^2)}    \\
  p_3 & = & \frac{2(2+2\kappa -\kappa^2)}{3(1+\kappa+\kappa^2)}  \nonumber
\end{eqnarray}
The range of $\kappa$ is then $ 0 \leq \kappa < +\infty$. In the
limit $\Lambda =0$ the Linet - Tian solution reduces to a form of
the Levi - Civita metric. In what follows it will be more useful to
change the coordinate $\rho$ to a new coordinate $x$, such that,
\begin{equation}\label{LTeq6}
    \sinh\left(\frac{\sqrt{3\Lambda}}{2} \rho\right) =  x
\end{equation}
so that the range of $x$ is also $0\leq x< \infty$. We then have,
\begin{eqnarray}\label{LTeq8}
   Q\left(\rho(x)\right) & = & \frac{2 x\sqrt{1+ x^2}}{\sqrt{3 \Lambda}} \nonumber \\
   P\left(\rho(x)\right) & = & \frac{2x}{\sqrt{3 \Lambda}\sqrt{1+ x^2}}
\end{eqnarray}
and the Linet - Tian metric takes the form,
\begin{eqnarray}\label{LTeq10}
ds^2 & = & -x^{\frac{2}{3}+p_1}\left(1+
x^2\right)^{\frac{1}{3}-\frac{p_1}{2}} dt^2
+x^{\frac{2}{3}+p_2}\left(1+ x^2\right)^{\frac{1}{3}-\frac{p_2}{2}} dz^2 \nonumber \\
& &+x^{\frac{2}{3}+p_3}\left(1+
x^2\right)^{\frac{1}{3}-\frac{p_3}{2}} d\phi^2 +\frac{4}{3\Lambda(1+
x^2)} dx^2
\end{eqnarray}
where, for simplicity and without loss of generality for the
analysis to be carried out in this paper, we have rescaled the
coordinates $t,z, \phi$. The metric (\ref{LTeq10}), up to rescalings
of the coordinates $t,z, \phi$, is the general solution of the
static Einstein equations with a negative cosmological constant,
with (full) cylindrical symmetry. As already indicated, the purpose
of this paper is to extend the stability analysis of the Levi -
Civita metric, given in \cite{glei1} to the Linet - Tian solution
with negative cosmological constant. The problem we are interested
in here is that of the (linear) evolution of perturbations of
(\ref{LTeq10}) that admit initial data of compact support, and that
break the cylindrical symmetry of the background, but preserve the
axial symmetry, so that we may write the perturbed metric in the
form,
\begin{equation}\label{gpeq2}
ds^2 = g^{(0)}_{\mu\nu}\left(1+\epsilon h_{\mu\nu}\right) dx^{\mu}
dx^{\nu}
\end{equation}
where $g^{(0)}_{\mu\nu}$ is the Linet - Tian metric (\ref{LTeq10}),
$x^{\mu}=\{t,x,z,\phi\}$, $h_{\mu\nu}=h_{\mu\nu}(t,x,z)$, and
linearity implies that all geometric quantities are computed up to
first order in $\epsilon$. From the linearity of the problem, and
the fact that $g^{(0)}_{\mu\nu}$ depends only on $x$, and leaving
aside for the moment the question of boundary conditions, we will
look first for solutions of the perturbation equations that have the
form,
\begin{equation}\label{gpeq4}
  h_{\mu\nu}\left(t,x,z\right) =e^{i(\Omega t - k z)}  H_{\mu\nu}(\Omega,k,x)
\end{equation}
We expect to find a ``complete set'' of such solutions, in the sense
that the evolution of arbitrary initial data may be (formally)
written as a Fourier transform of the form,
\begin{equation}\label{gpeq4}
  {h}_{\mu\nu}\left(t,x,z\right) =\int e^{i(\Omega t - k z)} {\cal{C}}(\Omega,k) H_{\mu\nu}(\Omega,k,x) d\Omega dk
\end{equation}
where ${\cal{C}}(\Omega,k)$ is determined by the initial data, and we must keep in mind that in the case of linear instabilities
some the $\Omega$ may be complex numbers, and, also, that part or
the whole spectrum of $\Omega$ may be discrete. On this account, we
write the perturbed metric in the form,
\begin{eqnarray}
\label{LT02}
  ds^2 &=& g^{(0)}_{tt}\left(1+\epsilon e^{i(\Omega t - k z)}
   F_1\right) dt^2+g^{(0)}_{xx}\left(1+\epsilon e^{i(\Omega t - k z)} F_2\right) dx^2 \nonumber \\
   & & +g^{(0)}_{zz}\left(1+\epsilon e^{i(\Omega t - k z)} F_3\right) dz^2
   +g^{(0)}_{\phi\phi}\left(1+\epsilon e^{i(\Omega t - k z)} F_4\right) d \phi^2 \nonumber\\
   & &  +2\epsilon e^{i(\Omega t - k z)} F_5 dt dx+2\epsilon e^{i(\Omega t - k z)} F_6 dt dz
   +2\epsilon e^{i(\Omega t - k z)} F_7 dx dz\\
    & & +2\epsilon e^{i(\Omega t - k z)} H_1 dt d\phi+2\epsilon e^{i(\Omega t - k z)} H_2 dx d\phi
   +2\epsilon e^{i(\Omega t - k z)} H_3 dz d\phi \nonumber
\end{eqnarray}
where the $F_i$ and $H_i$ are functions of $x$ only.

If we compute Einstein's equations using the form (\ref{LT02}) of
the metric, and retain only first order in $\epsilon$, we find that
the $F_i$ satisfy a set of coupled linear O.D.E., decoupled from the
$H_i$, which themselves satisfy a separate set of coupled linear
O.D.E. We may then consider as separate problems the cases where all
the $F_i$ are set equal to zero, (which we will call case I), and
the cases where only those terms are non vanishing (case II).

The set of perturbation equations that result from computing
Einstein's equations  are only {\em local}. In trying to extend
their solutions to the whole space time we are confronted with the
fact that the Linet - Tian solution is not globally hyperbolic. This
non global hyperbolicity stems from two facts. The first is that the
space time contains a time like singularity for $x=0$. This
singularity is the same as that present in the Levi - Civita metric.
The problem of analyzing the evolution of perturbations under this
condition was studied in \cite{glei1}. There it was shown that if
one imposes certain physically acceptable restrictions as a boundary
condition for $x=0$, it is possible to define a unique evolution for
arbitrary perturbations that satisfy the imposed conditions. As a
consequence it was possible to prove that the Levi-Civita space time
is unstable under perturbations that break the translational symmetry
along the symmetry axis. We may apply the same criterion here, but
in the Linet - Tian case we have another source of non global
hyperbolicity, because the boundary $x \to \infty$ is time like.
This has as a consequence that the constant $t$ hypersurfaces are
not Cauchy surfaces, and that there are null geodesics that remain
to the future of any given constant $t$ hypersurface, and never
intersect the hypersurface. Thus, the future evolution of any
perturbation may be arbitrarily modified by information incoming
from $x=+\infty$. One way out of this problem is to impose boundary
conditions, possible with some physical justification that
supplements the model, such that the evolution is solely determined
by initial data on some constant $t$ hypersurface. We notice,
however, that this is not a unique prescription, and that the
resulting evolution turns to be eventually dependent on the boundary
condition. We will therefore rephrase our problem, and try to find,
for a certain family of possible boundary conditions, under what
conditions, if any, we obtain a well defined, although possibly
unstable, evolution of appropriate initial perturbative data. These
and other points will be clarified as we derive our may results. We
refer to \cite{wald} for a more detailed discussion of this problem.

The plan of the paper is as follows. In the next Section we analyze
the gauge problem, with the purpose of extracting appropriate gauge
invariant quantities, and show that, as already indicated, the
evolution equations for a general perturbation separate into two
independent sets, that we call Case I and Case II. In Section III we
analyze Case I. We derive a ``master'' equation and obtain the
conditions under which there are unstable modes associated to the
evolution of type of perturbation. Case II is analyzed in Section
IV. This is the main part of the paper. We show that the
perturbations can be reduced to a diagonal form and we obtain a
``master'' variable, such that solving its equation of motion one
obtains a complete solution for the full perturbation. Nevertheless
this function is not gauge invariant, but we show how to extract the
corresponding gauge invariant part. Next we show that the solutions
of the equation satisfied by this gauge invariant part are in one -
to - one correspondence with the solutions of the eigenvalue -
eigenfunction problem for an operator in a one dimensional, single
particle, Schr\"odinger like system. The self adjointness of this
operator requires imposing appropriate boundary conditions that are
analyzed in detail. The structure of the resulting spectrum for the
operator is studied in Section V, where we find that essentially in
all cases it contains negative eigenvalues which correspond to
imaginary frequencies, thus establishing our main result that the
Linet - Tian space times are gravitationally unstable. The limit $k=0$, corresponding to purely radial perturbations is analyzed in Section VI. An interesting and somewhat unexpected result of this analysis is that, contrary to what happens for the Levi - Civita case, the Linet - Tian space times with negative cosmological constant are also {\em unstable} under purely radial perturbations. In Section VII we consider the limit $\kappa=0$, corresponding to the Bonnor metric, and find that it is {\em stable} under the linear perturbations considered here. We add, mostly for completeness, a brief comment on the ``hoop conjecture'' in Section VIII
Some final
comments are given in Section IX.

\section{The gauge problem.}

A central problem in the analysis of the evolution equations (as in
\cite{glei1}) is the elucidation of the gauge invariance of the
perturbations. We consider this problem in this Section. We first
notice that a general coordinate transformation from $(t,x,z,\phi)$
to new coordinates $(T,X,Z,\Phi)$ that preserves the form
(\ref{LT02}) of the perturbed metric may be written as,
\begin{eqnarray}
\label{LT04}
   t &=&  T + \epsilon e^{i(\Omega T - k Z)}T_1(X)\nonumber \\
     x &=&  X + \epsilon e^{i(\Omega T - k Z)}X_1(X)  \\
     z &=&  Z + \epsilon e^{i(\Omega T - k Z)}Z_1(X)\nonumber \\
      \phi &=&  \Phi + \epsilon e^{i(\Omega T - k Z)} \Phi_1(X) \nonumber
\end{eqnarray}

Indicating with a tilde the transformed coefficients, under this
transformation we have,
\begin{eqnarray}
\label{LT06}
   \widetilde{F}_1(X) &=& F_1(X)+  2 i \Omega T_1(X) +\frac{(3 p_1+4 X^2+2)  }{3X(1+ X^2)}X_1(X) \nonumber \\
   \widetilde{F}_2(X) &=& F_2(X)-\frac{2 X}{1+ X^2} X_1(X) +2 \frac{d X_1(X)}{dX}  \nonumber \\
   \widetilde{F}_3(X) &=& F_3(X)  - 2 i k Z_1(X) +\frac{(3 p_2+4 X^2+2)  }{3X(1+ X^2)}X_1(X)\nonumber \\
   \widetilde{F}_4(X) &=& F_4(X)+ \frac{(4 x^2+2 +3p_3)}{3 X(1+ x^2)}X_1(X)     \\
   \widetilde{F}_5(X) &=& F_5(X)+   \frac{4 i \Omega}{3\Lambda(1+ X^2)} X_1(X)- X^{p_1+2/3}(1+ X^2)^{1/3-p_1/2}\frac{d T_1(X)}{dX}  \nonumber \\
   \widetilde{F}_6(X) &=& F_6(X)+ \frac{i k X^{p_1+2/3}}{(1+ X^2)^{p_1/2-1/3}} T_1(X)
                   +\frac{i \Omega X^{p_2+2/3}}{(1+ X^2)^{p_2/2-1/3}} Z_1(X) \nonumber \\
   \widetilde{F}_7(X) &=& F_7(X)-  \frac{4i k}{3 \Lambda(1+ X^2)} X_1(X) + \frac{X^{p_2+2/3}}{(1+ X^2)^{p_2/2-1/3}} \frac{d Z_1(X)}{dX}
    \nonumber
\end{eqnarray}
and,
\begin{eqnarray}
\label{LT08}
   \widetilde{H}_1(X) &=& H_1(X)+  \frac{i \Omega X^{2/3+p_3}}{ (1+ X^2)^{p_3/2-1/3}} \Phi_1(X)    \nonumber \\
   \widetilde{H}_2(X) &=& H_2(X)+ \frac{X^{2/3+p_3}}{ (1+X^2)^{p_3/2-1/3}} \frac{d\Phi_1(X)}{dX}  \\
   \widetilde{H}_3(X) &=& H_3(X) - \frac{i k X^{2/3+p_3}}{ (1+ X^2)^{p_3/2-1/3}} \Phi_1(X) \nonumber
\end{eqnarray}
so that the $F_i$ and the $H_i$ transform separately, which is, of
course, consistent with the fact that they satisfy separate systems
of coupled O.D.E. We will consider the consequences of these forms
of the transformations on the gauge issues and the resulting
equations of motion in the following Sections.

\section{Case I.}

 In this Section we consider Case I. Here, in accordance with (\ref{LT08}) we
only have one free function ($\Phi_1(X) $) at our disposal, so we
may choose a gauge where any one, (or some particular combination)
of the functions $H_i$ is set equal to zero, and this removes any
gauge ambiguity. In more detail, it should be clear, from
(\ref{LT08}), that the expressions,
\begin{eqnarray}
\label{casII01}
   {\cal{G}}_1(x) &=&   k H_1(x)+\Omega H_3(x) \nonumber \\
   {\cal{G}}_2(x) &=&   H_2(x)+\frac{i}{\Omega}\frac{d H_1}{dx} -\frac{i (4 x^2+2 +3p_3)}{3\Omega x (1+ x^2)} H_1
\end{eqnarray}
are gauge invariant. On the other hand, in accordance with our
previous discussion, without loss of generality, we may choose a
gauge where $H_1(x)=0$. In this case (i.e. if we set $H_1=0$) the
functions $H_2$ and $H_3$ correspond to gauge invariants. We may,
therefore consider the perturbation equations that result when we
set $H_1(x)=0$ from the start. A simple computation then shows that
Einstein's equations imply that $H_2$ and $H_3$ must satisfy the
coupled set of ODEs,
\begin{eqnarray}
\label{casII04}
  \frac{d H_2}{dx} &=&  \frac{(3 p_1-1 -5 x^2)}{3 x (1+x^2)} H_2+\frac{4 i k x^{-p_2-2/3}}{(1+x^2)^{4/3-p_2/2}}H_3 \nonumber \\
  \frac{d H_3}{dx} &=& \frac{i}{k}\left[ \frac{ \Omega^2 x^{p_2-p_1}}{(1+ x^2)^{p_2/2-p_1/2}}- k^2 \right] H_2 + \frac{(2+4 x^2+3 p_3)}{3 x(1+ x^2)}H_3
\end{eqnarray}
We may use the first of these to solve for $H_3$ in terms of $H_2$
and $dH_2/dx$. Replacing in the second we find a second order
equation for $H_2$ that can be written in the form,
\begin{eqnarray}
\label{casII08}
  \frac{d^2 H_2}{dx^2} &=&  -\frac{(6 p_2+1 +11 x^2)}{3 x (1+ x^2)} \frac{dH_2}{dx}
       -\frac{4  \Omega^2 (1+ x^2)^{p_1/2-4/3}}{3 \Lambda x^{2/3+p_1} }H_2 \nonumber \\
    & & + H_2\left[\frac{4 k^2 (1+ x^2)^{p_2/2-4/3}}{3 \Lambda x^{2/3+p_2}} \right.
  \\
   & & \left.+ \frac{5 x^4 +(10 p_2+8 p_1+4)x^2+(1-3 p_1)(p_1+2p_2-1)}{3 x^2(1+ x^2)^2} \right] \nonumber
\end{eqnarray}

We immediately notice in this equation that $\Lambda$
appears only combined with $k$ and $\Omega$ in the forms
$k^2/\Lambda$ and $\Omega^2/\Lambda$. We may therefore define two
new parameters $\widetilde{k}=k/\sqrt{\Lambda}$, and
$\widetilde{\Omega}=\Omega/\sqrt{\Lambda}$, and the solutions of
(\ref{casII08}), for fixed $\kappa$, will be parameterized by
$\widetilde{k}$, and $\widetilde{\Omega}$, which is (formally)
equivalent to setting $\Lambda=1$ in (\ref{casII08}). We shall adopt
this latter choice, remembering that for $\Lambda \neq 1$ we must
rescale the values of $k$ and $\Omega$.

Suppose now we fix the value of $\kappa$, which fixes the
unperturbed space time. In principle, we may specify $k$ freely.
Then, if we impose boundary conditions on (\ref{casII04}) for $x=0$,
and $x \to \infty$, we turn the problem of solving the equation into
a boundary value problem that determines the allowed values of
$\Omega^2$. In principle we would have an infinite set of such
solutions, that can be considered as functions of $k$, and $\Omega$,
and which, through the use (\ref{casII04}), provides a complete
solution of the perturbation equations for the given parameters
Then, plugging these solutions in (\ref{gpeq4}) we would have an
infinite set of solutions of the evolution equations for the
perturbations.

We are now confronted with several problems. One is that we do not
have closed form solutions of this equation. Another is that it is
not clear how to impose acceptable boundary conditions that define
and restrict the range of possible values of $\Omega$. And, most
importantly, how can we assure that our set of solutions is
``complete'', in the sense that it can, at least in principle,
describe the evolution of acceptable but arbitrary initial data.

These problems may be analyzed by constructing an equivalent self
adjoint operator, whose eigenvalues and eigenfunctions are in one to
one correspondence with the solutions of (\ref{gpeq4}), and the
corresponding values of $\Omega$. This can be achieved by
introducing a new function $H_4(y)$, where $y$ is a function of $x$,
and an ``integration'' function ${\cal{K}}(x)$ such that,
\begin{equation}\label{casII10}
    H_2(x)={\cal{K}}(x) H_4(y(x))
\end{equation}
where $y(x)$ satisfies the equation,
\begin{equation}\label{casII12}
    \frac{dy}{dx} = \frac{2}{\sqrt{3} x^{p_1/2+1/3}(1+ x^2)^{2/3-p_1/4}}
\end{equation}
and,
\begin{equation}\label{casII14}
    {\cal{K}}(x)=x^{p_1/4-p_2}(1+x^2)^{p_2/2-p_1/8-1/2}
\end{equation}

Replacing in (\ref{casII08}) we find that $H_4(y)$ satisfies the
equation,
\begin{equation}\label{casII16}
  - \frac{d^2 H_4}{dy^2} +{\cal{V}}_{II}(y) H_4 = \Omega^2 H_4
\end{equation}
where,
\begin{eqnarray}\label{casII18}
   {\cal{V}}_{II}(y) & = & \frac{x^{p_1-p_2} k^2}{(4+3 \Lambda x^2)^{p_1/2-p_2/2}} \nonumber \\
   & & +\frac{ (9 p_1+12 p_2-4)(5 p_1+4 p_2 -4) -16 (7 p_1+8 p_2) x^2  }{64 x^{4/3-p_1}(1+ x^2)^{2/3+p_1/2}}
\end{eqnarray}
and $x$ should be given as a function of $y$ by inverting $y(x)$. We
recognize that (\ref{casII16}) has the standard form of the
Schr\"odinger equation for a particle in a one dimensional
potential, and, therefore, its spectrum can analyzed  using well
known standard procedures. In particular, its self adjoint
extensions provide a complete, orthonormal, basis of functions in
its domain.

The expression (\ref{casII18}) for $ {\cal{V}}_{II}$ is not the most
useful because $p_1$ and $p_2$ are not independent. A more useful
one in terms of $\kappa$ is obtained using (\ref{LTeq5}),
\begin{eqnarray}\label{casII28}
   {\cal{V}}_{II}(y) & = & \frac{x^{\frac{2 \kappa (2 +\kappa)}{1 +\kappa+\kappa^2}} k^2}
   {\Lambda (1+ x^2)^{\frac{ \kappa (2 +\kappa)}{1 +\kappa+\kappa^2}}} \nonumber \\
   & & +\frac{3(4 \kappa+5)(4\kappa+3)+8 (1 +\kappa+\kappa^2)  (5 +6 \kappa -2 \kappa^2)  x^2}
   {16(1 +\kappa+\kappa^2)^2 x^{\frac{2}{1 +\kappa+\kappa^2}}(1+ x^2)^{\frac{(2 \kappa+1)^2}{3(1 +\kappa+\kappa^2)}}}
\end{eqnarray}
We have, therefore, that for $x \to 0$ the potential diverges to
$+\infty$, while for $x \to \infty$ we have ${\cal{V}}_{II} \to k^2/
\Lambda$.  We further notice that for $2 \kappa^2 \leq 5 + 6 \kappa$
(i.e. for $\kappa \leq (3+\sqrt{19})/2 = 3.67...)$, the potential
${\cal{V}}_{II}$ is positive definite, and, therefore, we will have
$\Omega^2\geq 0$ for any self adjoint extension of (\ref{casII16}).
On the other hand, for $\kappa$ larger than this value, there are
negative terms in the potential, and these, in turn, might dominate
in some region, making ${\cal{V}}_{II}$ negative there. To see what
effect this may have on the allowed values of $\Omega$ we need to
establish the existence of self adjoint extensions for
(\ref{casII16}), and this requires making the dependence of $
{\cal{V}}_{II}(y)$ on $y$ more explicit.

In more detail, we may fix an integration constant in (\ref{casII12})
and set,
\begin{equation}\label{casII20}
   y(x) = \int_0^x{\frac{2}{\sqrt{3}x_1^{p_1/2+1/3}(1+x_1^2)^{2/3-p_1/4}} dx_1}
\end{equation}
Since,
\begin{equation}\label{casII22}
   \frac{p_1}{2}+\frac{1}{3} = \frac{\kappa+\kappa^2}{1+\kappa+\kappa^2}
\end{equation}
the integral in (\ref{casII20}) converges for all $x$, and we have
$0 \leq y \leq y_0$ for $0 \leq x < \infty$, with,
\begin{equation}\label{casII20a}
   y_0 = \int_0^{\infty}{\frac{2}{\sqrt{3}x_1^{p_1/2+1/3}(1+ x_1^2)^{2/3-p_1/4}} dx_1}
\end{equation}
We also notice that near $x=0$ we have (to leading orders) the
expansions,
\begin{eqnarray}
\label{casII24}
y(x) &=& \frac{4\sqrt{3} x^{2/3-p_1/2}}{4+3 p_1 } \left[1 +\frac{(3p_1-8)(4+3 p_1) x^2}{12(16-3p_1)}\right]  \nonumber \\
x(y) &=& \left(\frac{(4-3p_1)y}{4
\sqrt{3}}\right)^{6/(4-3p_1)}+\frac{
(3p_1-8)}{2(3p_1-16)}\left(\frac{(4-3p_1)y}{4
\sqrt{3}}\right)^{18/(4-3p_1)}
\end{eqnarray}
while for $x \to \infty$ we have, to leading order,
\begin{eqnarray}
\label{casII26}
y(x) &=&  y_0 -\frac{\sqrt{3}}{x^{2/3}}\nonumber \\
x(y) &=& \frac{\sqrt{3}}{ (y_0-y)^{3/2}}
\end{eqnarray}
\\

Using these results, we first notice that from (\ref{casII24}), near
$y=0$, to leading order, we have,
\begin{equation}\label{casII30}
  {\cal{V}}_{II}(y) \simeq \frac{(5+4 \kappa)(3+4 \kappa)}{4 y^2}
\end{equation}
and, therefore, near $y=0$, the general solution of (\ref{casII16})
behaves as,
\begin{equation}\label{casII34}
  H_4(y) \sim C_1 y^{2 \kappa +5/2} +C_2 y^{-2 \kappa-3/2}
\end{equation}
where $C_1$ and $C_2$ are arbitrary constants. This implies that the
second term on the RHS of (\ref{casII34}) diverges faster than
$1/y$, and, therefore, we must set $C_2=0$ to have a boundary
condition appropriate for a self adjoint extension.

The other important limit is $ y \to y_0$ ($x \to \infty$). In this
limit we have
\begin{equation}\label{casII36}
  {\cal{V}}_{II}(y) \simeq  \frac{k^2}{ \Lambda }
  -\frac{(5 +6 \kappa-2\kappa^2)}{4 \sqrt{3}} (y_0-y)
\end{equation}
and, therefore, the general solution of (\ref{casII16}), for $ y \to
y_0$ is of the form,
\begin{equation}\label{casII38}
  H_4(y) \sim C_3  +C_4 (y_0-y)
\end{equation}
where $C_3$ and $C_4$ are arbitrary constants. As regards the
possibility of a self adjoint extension this corresponds to the {\em
circle limit} case. In this case we obtain a self adjoint extension
by imposing that all allowed solutions must satisfy the boundary
condition
\begin{equation}\label{casII40}
  C_3  = \alpha C_4
\end{equation}
for some {\em fixed} $\alpha$. This implies that there are infinite
possible different self adjoint extensions. In particular for
$\alpha=0$ ($C_3=0$) we have the Dirichlet and for $\alpha=\infty$
($C_4=0$) the Neumann boundary condition. The general case of
arbitrary $\alpha$ is called the Robin boundary condition. This
large ambiguity in the boundary condition for $y \to y_0$  can be
traced to the fact, already discussed, that for adS the region $x
\to \infty$ corresponds to a time like boundary.

We now go back to the problem of finding the allowed $\Omega$,
assuming we have chosen a particular self adjoint extension. As
remarked, for $2 \kappa^2 < 5 +6 \kappa$, the potential is positive
definite, and, therefore, $\Omega^2 >0$ for all self adjoint
extensions, corresponding to a stable evolution of arbitrary initial
data. On the other hand for sufficiently large $\kappa$ the
potential is negative in some region, as can be explicitly checked
by plotting $ {\cal{V}}_{II}(y)$ as a function of $y$, using
(\ref{casII28}), and (\ref{casII20}) as parametric representations
of these functions. An example is given in Fig. 1. It can be seen
that the region ${\cal{V}}_{II}(y) <0 $ is comparatively small, and
shallow. We have analyzed numerically the possible existence of
solutions with $\Omega^2<0$ and found that even for large values of
$\kappa$ these are possible only for $\alpha > 0$. For instance, for
the example of Fig. 1, we find that there are solutions with
$\Omega^2 < 0$ only for $\alpha > 1.29...$. The existence of these
solutions can be understood if we consider solving numerically
(\ref{casII16}), by imposing the appropriate boundary condition at
$y=0$, for some negative value of $\Omega^2$, and integrating
towards $y =y_0$. Since the potential is smooth for $y > 0$, we
expect to find a smooth solution for $y > 0$, such that for $y =
y_0$ we have some finite values for the solution and its first
derivative, and, therefore, some finite value of $\alpha$. We shall
not elaborate further on these type of solutions and remark only
that there appear to be no solutions for $\Omega^2<0$ if we impose
the Dirichlet boundary condition for $y = y_0$, and, therefore, the
Case I perturbations are stable with respect to that boundary
condition. In the next Section we consider Case II.

\begin{figure}
\centerline{\includegraphics[height=12cm,angle=-90]{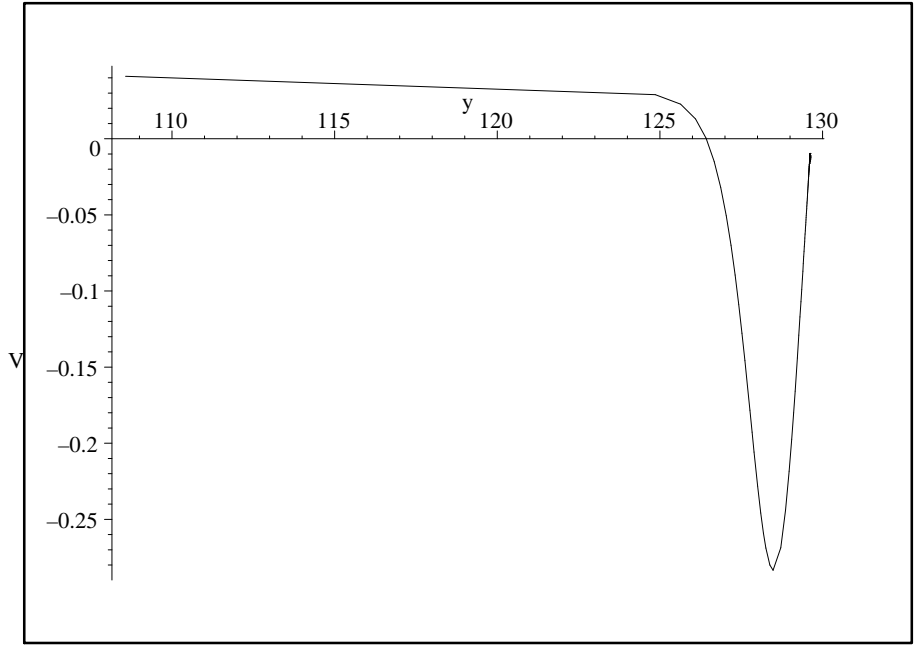}}
\caption{${\cal{V}}_{II}(y)$ as a function of $y$ for $\kappa=10$
and $k=0.01$. Notice that only the region where ${\cal{V}}_{II}(y)
<0 $ is shown. The curve extends to the left, with positive values,
up to $y=0$ where it diverges to $+\infty$.}
\end{figure}

\section{Case II.}

Consider now Case II. In this case, in accordance with (\ref{LT06}),
given an arbitrary perturbation, since the  functions  $T_1$, $X_1$,
and $Z_1$ are arbitrary, we may choose them such that
$\widetilde{F}_1 =0$, $\widetilde{F}_3 =0$, and, $\widetilde{F}_4
=0$. It should be clear that with this choice we remove all gauge
ambiguity, because any further transformation would make at least
one these functions different from zero. If we make this choice, and
assume that only $F_2(x)$, $F_5(x) $, $F_6(x)$, and, $F_7(x)$ are
non zero, replacing in the linearized Einstein equations we obtain a
coupled set of O.D.E.s for these functions such that by appropriate
replacements we may express explicitly  $F_5(x) $, $F_6(x)$, and,
$F_7(x)$, as linear functions of $F_2$ and $dF_2/dx$, while $F_2$,
in turn, satisfies a second order O.D.E. Unfortunately, the explicit
forms of the coefficients that result, because of their length and
complexity, are very difficult to handle. For this reason, we have
chosen to analyze a different choice of gauge, that, although not
free of gauge ambiguities, is much easier to handle, and eventually
leads to gauge invariant results. In more detail, since the
functions $T_1$, $X_1$, and $Z_1$ are arbitrary, it is clear that we
can always choose them such that
$\widetilde{F}_5=\widetilde{F}_6=\widetilde{F}_7=0$. Thus, without
loss of generality, we may restrict case II to a ``diagonal form'',
namely, we assume that only  $F_1$, $F_2$, $F_3$, and $F_4$, are non
vanishing. Once this choice is made, there appears to be no freedom
left for further simplifications. It turns out, however, that the
``diagonal form'' is not gauge invariant. This is because the
system,
\begin{eqnarray}
\label{LT10}
   0 &=&  \frac{4 i \Omega}{4+3\Lambda X^2} X_1(X)- X^{p_1+2/3}(4+3\Lambda X^2)^{1/3-p_1/2}\frac{d T_1(X)}{dX}  \nonumber \\
    0 &=&  \frac{i k X^{p_1+2/3}}{(4+3\Lambda X^2)^{p_1/2-1/3}} T_1(X) +\frac{i \Omega X^{p_2+2/3}}{(4+3\Lambda X^2)^{p_2/2-1/3}} Z_1(X)  \\
   0 &=& -  \frac{4i k}{4+3\Lambda X^2} X_1(X) + \frac{X^{p_2+2/3}}{(4+3\Lambda X^2)^{p_2/2-1/3}} \frac{d Z_1(X)}{dX}      \nonumber
\end{eqnarray}
has the solution,
\begin{eqnarray}
\label{LT12}
   T_1(X) &=&  -2 i \Omega Q_1 X^{p_2/2-p_1/2}\left(4+3 \Lambda X^2\right)^{p_1/4-p_2/4}  \nonumber \\
    X_1(X) &=& \left(p_1-p_2\right) Q_1 X^{p_2/2+p_1/2-1/3}\left(4+3 \Lambda X^2\right)^{1/3-p_1/4-p_2/4}  \\
   Z_1(X) &=& 2 i k Q_1 X^{p_1/2-p_2/2}\left(4+3 \Lambda X^2\right)^{p_2/4-p_1/4}  \nonumber
\end{eqnarray}
where $Q_1$ is an arbitrary function of $k$ and $\Omega$. This
implies, as can be easily checked, that the system,
\begin{eqnarray}
\label{LT14}
   F_1(x) &=&  - Q_1 \left(\frac{16 \Omega^2  x^{p_2/2-p_1/2}}{3 \Lambda(1+x^2)^{p_2/4-p_1/4}}
    + \frac{(p_1-p_2)(3 p_1+4 x^2 +2) x^{p_1/2+p_2/2-4/3}}{3(1+ x^2)^{p_1/4+p_2/4+2/3}}\right)\nonumber \\
    F_2(x) &=&  \frac{  Q_1(p_1-p_2) (2 +4 x^2+3 p_3)(1+ x^2)^{p_3/4-2/3}}{x^{p_3/2+4/3}} \nonumber \\
    F_3(x) &=&  - Q_1 \left(\frac{16 k^2  x^{p_1/2-p_2/2}}{(1+ x^2)^{p_1/4-p_2/4}}
    + \frac{(p_1-p_2)(3 p_2+4 x^2 +2) x^{p_1/2+p_2/2-4/3}}{3(1+ x^2)^{p_1/4+p_2/4+2/3}}\right)\  \\
    F_4(x) &=& - \frac{  Q_1(p_1-p_2) (2 +4 x^2+3 p_3)(1+ x^2)^{p_3/4-2/3}}{x^{p_3/2+4/3}}  \nonumber
\end{eqnarray}
with $F_5(x)=F_6(x)=F_7(x)=0$, and $Q_1$ an arbitrary function of
$\Omega, k$, is a pure gauge solution of the perturbed Einstein
equations.
\\

Thus the ``diagonal'' perturbations are not gauge invariant, because
there are coordinate transformations that preserve the diagonal
form. If we indicate with $F_i(x)$ a particular representation and
with $\widetilde{F}_i(x)$ the transformed perturbation, in
accordance with (\ref{LT06}) and (\ref{LT12}) they are related by,
\begin{eqnarray}
\label{peq34}
  {\widetilde{F}}_1(x) &=&  F_1(x) +\frac{16 \Omega^2 Q_1 (1+ x^2)^{p_1/4-p_2/4}}{3 \Lambda (p_1-p_2)x^{p_1/2-p_2/2}}
                       +\frac{ Q_1(4 x^2+3 p_1+2)x^{p_1/2+p_2/2-4/3}}{3(1+ x^2)^{p_1/4+p_2/4+2/3}}     \nonumber \\
 {\widetilde{F}}_2(x) &=&  F_2(x) -\frac{ Q_1 (4 x^2+2 +3 p_3)x^{p_1/2+p_2/2-4/3}}{3 (1+x^2)^{p_1/4+p_2/4+2/3}}
                         \\
  {\widetilde{F}}_3(x) &=&  F_3(x) +\frac{16 k^2 Q_1 (1+ x^2)^{p_2/4-p_1/4}}{3(p_1-p_2)x^{p_2/2-p_1/2}}
                       +\frac{ Q_1(4 x^2+3 p_2+2)x^{p_1/2+p_2/2-4/3}}{3(1+ x^2)^{p_1/4+p_2/4+2/3}}     \nonumber \\
  {\widetilde{F}}_4(x) &=&  F_4(x) +\frac{ Q_1 4 x^2+2+ 3 p_3)x^{p_1/2+p_2/2-4/3}}{ 3(1+ x^2)^{p_1/4+p_2/4+2/3}}
                          \nonumber\\
\end{eqnarray}
This immediately implies that, although the functions $F_i$ are not
gauge invariant, we have that
\begin{equation}
\label{peq36}
 {\widetilde{F}}_2(x)+{\widetilde{F}}_4(x)=   F_2(x)+ F_4(x)
\end{equation}
and, therefore, $F_2(x)+ F_4(x)$ is a gauge invariant quantity.
Similarly,
\begin{eqnarray}
\label{peq38}
  {\cal{G}}_{14}(x) &=&  F_1(x) - \frac{16 (1+ x^2)^{p_1/2+2/3} \Omega^2 F_4(x)}
  {\Lambda (p_1-p_2)x^{p_1-4/3}(4 x^2+2 +3 p_3)}
                    -\frac{(4x^2+3 p_1+2) F_4(x)}{4 x^2+2 +3 p_3)}  \nonumber \\
  {\cal{G}}_{34}(x) &=&  F_3(x) - \frac{16 (1+ x^2)^{p_2/2+2/3} k^2 F_4(x)}
  {\Lambda(p_1-p_2)x^{p_2-4/3}(4 x^2+2 +3 p_3)}
                    -\frac{(4 x^2+3 p_2+2) F_4(x)}{4x^2+2 +3 p_3)}
\end{eqnarray}
are also gauge invariant quantities.

Replacing the diagonal form of the perturbations in Einstein's
equations, and expanding to first order in $\epsilon$, we get a set
of coupled second order ordinary differential equations for the
functions $F_i(x)$. A closer examination shows that for $\Omega\neq
0$, and $z\neq 0$, we have,
\begin{equation}\label{peq4}
    F_2(x)=-F_4(x)
\end{equation}
while for $F_1$ and $F_3$ we find,
\begin{eqnarray}
\label{peq6}
  \frac{dF_1}{dx} &=& -\frac{dF_4}{dx} - \frac{(3 p_1-3 p_2+6p_3 +8 x^2+4)}
  {6 x (1+x^2)} F_4 -\frac{(p_1-p_2)}{2 x(1+ x^2)} F_1 \nonumber \\
 \frac{dF_3}{dx} &=& -\frac{dF_4}{dx} +
  \frac{(3 p_1-3 p_2-6p_3 -8 x^2-4)}{6 x (1+x^2)} F_4 +\frac{(p_1-p_2)}{2 x(1+ x^2)} F_3
\end{eqnarray}

We also find,
\begin{equation}
\label{peq8}
  \frac{dF_4}{dx} = {\cal{R}}_1 F_1+{\cal{R}}_2 F_2+{\cal{R}}_3 F_3
\end{equation}
where,
\begin{eqnarray}
\label{app03}
 {\cal{R}}_1  &=&  -\frac{ (8 x^2-3 p_1 +4)(p_1-p_2)\Lambda +16 k^2 x^{4/3-p_2}(1+x^2)^{2/3+p_2/2}}
                      {\Lambda x (4 x^2+3p_3+2)(1+x^2)} \nonumber \\
 {\cal{R}}_2  &=& \frac{ (8 x^2-3 p_2 +4)(p_1-p_2)\Lambda +16 \Omega^2 x^{4/3-p_1}(1+x^2)^{2/3+p_1/2}}
                      {\Lambda x (4 x^2+3p_3+2)(1+x^2)}  \\
{\cal{R}}_3  &=& \frac{32 x^4 +(120 p_3+32)x^2 +45 p_1 p_2 +44+60
p_3}{12 x (1+x^2)(4 x^2+3p_3+2)} \nonumber \\
      & & -\frac{4 (1+x^2)^{p_1/2-1/3} x^{1/3-p_1} \Omega^2}{\Lambda
      (4 x^2+3p_3+2)} + \frac{4 (1+x^2)^{p_2/2-1/3} x^{1/3-p_2} k^2}{\Lambda
      (4 x^2+3p_3+2)} \nonumber
\end{eqnarray}

One can now show that any set of functions $(F_1, F_2, F_3, F_4)$
that satisfy (\ref{peq4},\ref{peq6},\ref{peq8}) is a solution of the
perturbation equations, and therefore, to solve those equations, all
we need to do is to solve the system
(\ref{peq4},\ref{peq6},\ref{peq8}). A first step is to introduce two
new functions, $G_1(x)$ and $G_2(x)$, defined by,
\begin{eqnarray}
\label{peq10}
  G_1(x) &=& F_3(x)+F_4(x) \nonumber \\
   G_2(x) &=& F_3(x)-F_4(x)
\end{eqnarray}

Now, using (\ref{peq6}) it follows that,
\begin{equation}
\label{peq12}
  G_2 =\frac{6x(1+ x^2)}{2+4x^2+3 p_3}\frac{dG_1}{dx} + \frac{ 2 +4 x^2-6 p_1}{2+4x^2 +3 p_3} G_1
\end{equation}

Then, from the definitions (\ref{peq10}) we obtain,
\begin{eqnarray}
\label{peq14}
  F_3 &=& \frac{3 x (1+ x^2)}{2 +4 x^2+ 3 p_3)} \frac{dG_1}{dx} +\frac{8 x^2+4-9 p_1-3 p_2}{8 x^2 +4 +3 p_3} G_1 \nonumber \\
  F_4 &=& -\frac{3 x (1+ x^2)}{4 x^2+2+3 p_3} \frac{dG_1}{dx} +\frac{3 p_1-3p_2 }{8 x^2+4+6 p_3} G_1
\end{eqnarray}
and, therefore, we have explicit expressions for $F_3$, $F_4$, (and
$F_2=-F_4$) in terms of $G_1$ and $dG_1/dx$.

We also find an expression $F_1$ in terms of $G_1$, which has the
form,
\begin{equation}
\label{peq16}
  F_1 = {\cal{R}}_4 \frac{d^2G_1}{dx^2}+{\cal{R}}_5 \frac{dG_1}{dx}+{\cal{R}}_6 G_1
\end{equation}
where ${\cal{R}}_4$, ${\cal{R}}_5$, and ${\cal{R}}_6$ are again
lengthy and complicated functions of $x$ and the other parameters in
the problem. If we replace these expressions for the $F_i$ in the
perturbation equations, we find that all the equations are
satisfied, provided $G_1(x)$ is a solution of a {\em third order}
O.D.E. of the form,
\begin{equation}
\label{peq18}
 \frac{d^3 G_1}{dx^3}+{\cal{R}}_7 \frac{d^2 G_1}{dx^2}+{\cal{R}}_8 \frac{d G_1}{dx}+{\cal{R}}_9  G_1 =0
\end{equation}
where the functions ${\cal{R}}_i$ depend on $x$, the metric
parameters, $k$ and $\Omega$ in a complicated way. Thus the problem
of analyzing the perturbations considered here is reduced to that of
finding the solutions of Eq. (\ref{peq18}). But, because of the
complicated form of the functions ${\cal{R}}_i$, this turns out be a
rather difficult task. We may, nevertheless, apply the previous
results on gauge invariance, to achieve a certain simplification of
the problem and take it to a form more suitable for its analysis. We
notice that under a coordinate transformation of the form
(\ref{LT12}) we have,
\begin{equation}
\label{peq40}
 {\widetilde{G}}_1(x) =  G_1(x) +C W_1(x)
\end{equation}
where $C$ is a constant, and,
\begin{equation}
\label{peq42}
 W_1(x) =  \frac{ 16  k^2 (1+ x^2)^{p_2/4-p_1/4}}{(p_1-p_2)x^{p_2/2-p_1/2}}
                        +\frac{\Lambda  (8 x^2+4 -3p_1) x^{p_2/2+p_1/2-4/3}}{(1+ x^2)^{p_1/4+p_2/4+2/3}}
\end{equation}

We remark that since $G_1(x)$ is already a first order quantity, the
{\em form} (and the coefficients) of (\ref{peq18}) are invariant
under the transformation (\ref{peq38}), and, therefore, {\em both}
$G_1(x)$ and ${\widetilde{G}}_1(x)$ must be solutions of
(\ref{peq18}), which implies, as can explicitly be checked, that
$W_1(x)$ is also a solution of (\ref{peq18}). In fact, it is a pure
gauge solution of (\ref{peq18}), and, therefore, corresponds to a
pure gauge solution of the perturbation equations. But, if we go
back to (\ref{peq40}), it is clear that,
\begin{equation}
\label{peq44}
 \frac{d}{dx}\left(\frac{{\widetilde{G}}_1(x)}{W_1(x)}\right) =  \frac{d}{dx}\left(\frac{G_1(x)}{W_1(x)}\right)
\end{equation}
and, therefore, $d(G_1/W_1)/dx$ is gauge invariant. If we write the
general solution of (\ref{peq18}) in the form,
\begin{equation}
\label{peq22}
  G_1(x)=W_1(x) W_2(x) + C W_1(x)
\end{equation}
replacing in (\ref{peq18}) we find that $W_2(x)$ satisfies an
equation of the form,
\begin{equation}
\label{peq24}
 \frac{d^3 W_2}{dx^3}+{\cal{R}}_{10} \frac{d^2 W_2}{dx^2}+{\cal{R}}_{11}\frac{d W_2}{dx} =0
\end{equation}
where, again, the functions ${\cal{R}}_i$ depend on $x$, the metric
parameters, $k$ and $\Omega$ in a complicated way. This implies that
the function,
\begin{equation}
\label{peq26}
 W_3(x) =\frac{d W_2}{dx}
\end{equation}
which, on account of (\ref{peq44}) is gauge invariant, and,
therefore, contains the non trivial part of the perturbation,
satisfies a {\em second order} O.D.E. The main difficulty in the
analysis of this equation is due to the presence in the coefficients
of (\ref{peq24}) of powers of $x$, (and of $(1+ x^2)$) that
are rational functions of $\kappa$ that do not combine in a simple
way. Nevertheless, we may show that in general the equation for
$W_3(x)$ may be written in the form,
\begin{equation}
\label{peq28}
 - \frac{\Lambda x^{p_1+2/3} \left(1+
x^2\right)^{4/3-p_1/2}}{4}\frac{d^2 W_3}{dx^2}+{\cal{R}}_{12}
\frac{d W_3}{dx}+{\cal{R}}_{13} W_3 =\Omega^2 W_3
\end{equation}
where the functions ${\cal{R}}_{12}$ and ${\cal{R}}_{13}$ depend on
$x$, $\kappa$, $\Lambda$, and $k$, but are independent of $\Omega$.
But, since we expect the solutions to satisfy some restrictions in
order to be acceptable as perturbations, for instance in their
behaviour for both $x \to 0$ and $x \to \infty$, it should be clear
that once these restrictions are imposed, (\ref{peq28}) may be taken
as a boundary value - eigenvalue problem that determines the acceptable
values of $\Omega$. We may then try applying known techniques, such
those used in \cite{glei1}, to find the allowed spectrum for
$\Omega$, and look for possible instabilities, signaled by a
negative value of $\Omega^2$. The first step would be to find
appropriate boundary conditions, such that the resulting functions
are acceptable as perturbations. However, because of the complicated
form of the equation satisfied by $W_3$, it turns out that it is
somewhat simpler to proceed directly to an analysis of its solutions
and then relate those solutions to the general problem. Just as in Case I, this can be
made to look like a quantum mechanical problem for a particle in a
one dimensional potential by introducing a new variable $y$, and a
new function, $W_4(y)$, such that,
\begin{equation}
\label{peq32} W_3(x)= {\cal{K}}(x) W_4\left(y(x)\right)
\end{equation}
where ${\cal{K}}(x)$ is an ``integration factor'', determined as
follows. We first write (\ref{peq28}) in the form,
\begin{equation}
\label{peq28a}
   \frac{d^2 W_3}{dx^2}+q_3(x)
\frac{d W_3}{dx}+q_4(x) W_3+ q_1(x)\Omega^2 W_3=0
\end{equation}
where $q_3$ and $q_4$ are related in an obvious way to
${\cal{R}}_{12}$ and ${\cal{R}}_{13}$ and
\begin{equation}
\label{peq28b}
 q_1(x) =  \frac{4}{3\Lambda x^{p_1+2/3} \left(1+
x^2\right)^{4/3-p_1/2}}
\end{equation}

Replacing (\ref{peq32}) in (\ref{peq28a}) we get,
\begin{eqnarray}
\label{peq28c}
&& - \frac{d^2W_4}{d^2 y}
-\left(\frac{dy}{dx}\right)^{-1}\left(\frac{2}{{\cal{K}}}\frac{d
  {\cal{K}}}{dx} +\left(\frac{dy}{dx}\right)^{-1}\frac{d^2 y}{dx^2} +q_3\right)\frac{d W_4}{d
  y} \nonumber \\
&&  -\frac{1}{{\cal{K}}} \left(\frac{dy}{dx}\right)^{-2} \left(
\frac{d^2 {\cal{K}}}{dx^2}+q_3 \frac{d
  {\cal{K}}}{dx}+q_4 {\cal{K}} \right) W_4 \\
&&  = \Omega^2 q_1 \left(\frac{dy}{dx}\right)^{-2} W_4 \nonumber
\end{eqnarray}

If we choose now $y(x)$ as a solution of,
\begin{eqnarray}
\label{peq30}
\frac{dy}{dx}& = & \sqrt{\Lambda q_1(x)} \nonumber \\
& = & \frac{2}{\sqrt{3}x^{p_1/2+1/3} \left(1+
x^2\right)^{2/3-p_1/4}}\;,
\end{eqnarray}
(which is identical to (\ref{casII12})), and impose the condition
that ${\cal{K}}$ is a solution of the equation,
\begin{equation}\label{peq28d}
    \frac{2}{{\cal{K}}}\frac{d
  {\cal{K}}}{dx} =-\left(\frac{dy}{dx}\right)^{-1}\frac{d^2 y}{dx^2}
  -q_3
\end{equation}
so that the coefficient of $dW_4/dy$ in (\ref{peq28c}) vanishes,
replacing in (\ref{peq28c}) we finally find that $W_4(y)$ satisfies
an  equation of the form,
\begin{equation}
\label{peq34} {\cal{H}} W_4 =\frac{\Omega^2}{\Lambda} W_4
\end{equation}
where,
\begin{equation}
\label{peq34a} {\cal{H}}= -  \frac{d^2 }{dy^2}+ {\cal{V}}_{1}(y)
\end{equation}
and
\begin{equation}
\label{peq34aa} {\cal{V}}_{1}(y) = -\frac{1}{{\cal{K}}}
\left(\frac{dy}{dx}\right)^{-2} \left( \frac{d^2
{\cal{K}}}{dx^2}+q_3 \frac{d
  {\cal{K}}}{dx}+q_4 {\cal{K}} \right)
\end{equation}
Actually, as will be clear in what follows, ${\cal{V}}_{1}$ is
readily obtained only as a function of $x$, but it is implicitly
dependent on $y$ through the function $x(y)$ obtained by inverting
$y(x)$. The main difficulty here is that, unfortunately, for general
values of $\kappa$ the function $q_3(x)$ is rather complicated and
we could not find explicit expressions for ${\cal{K}}(x)$.
Nevertheless one can check that with the choice (\ref{peq30}) of
$y(x)$ the  right hand side of (\ref{peq28d}) is regular in $0 < x <
\infty$, and therefore, the solution ${\cal{K}}(x)$ is also regular
in $0 < x < \infty$. Moreover, if we assume that ${\cal{K}}(x)$
satisfies (\ref{peq28d}), replacing in (\ref{peq34aa}) we obtain,
\begin{equation}
\label{peq34b}
 {\cal{V}}_{1}(y)=\frac{1}{4}\left(\frac{dy}{dx}\right)^{-4}\left(2 \frac{dy}{dx}\frac{d^3y}{dx^3}
 -3\left(\frac{d^2y}{dx^2}\right)^2\right)
 +\frac{1}{4}\left(\frac{dy}{dx}\right)^{-2}\left(q_3^2+2\frac{dq_3}{dx}-4q_4\right)
\end{equation}

Notice that an important consequence of this result is that, besides
$q_3$, $q_4$, and $dy/dx$ (given by (\ref{peq30})), {\em no} explicit
knowledge of ${\cal{K}}$, or of $y(x)$, is required to construct an
explicit form of $ {\cal{V}}_{1}$ as a function of $x$.  It is,
again, understood that $x$ should be given as function $y$ by
inverting (\ref{peq30}), but, as we will see in what follows, the x-dependent
form (\ref{peq34b}) will be all we need to obtain the general properties of $
{\cal{V}}_{1}(y)$, and, therefore, of the solutions of (\ref{peq34}).
One further important property that results from the explicit x-dependent form
of $ {\cal{V}}_{1}$ is that, besides $\kappa$, it depends only on
$k/\sqrt{\Lambda}$. Therefore, if we introduce,
\begin{eqnarray}
\label{Omktil}
\widetilde{\Omega} & = & \Omega/\sqrt{\Lambda} \nonumber \\
\widetilde{k} & = & k/\sqrt{\Lambda}\;,
\end{eqnarray}
then (\ref{peq34a}) takes the form
\begin{equation}
\label{peq34ab}
 -  \frac{d^2 W_4 }{dy^2}+ {\cal{V}}_{1}(\kappa,\widetilde{k},y) W_4
 = \widetilde{\Omega} W_4
\end{equation}
which contains {\em no} explicit reference to $\Lambda$. Thus, for
fixed $\kappa$, the effect of changing $\Lambda$ in the unperturbed
metric is just a rescaling of the parameters $k$, and $\Omega$ in
the solutions of the perturbation equations.

Going back to (\ref{peq34}), we notice that it has the familiar form
of a (one dimensional) Schr\"odinger equation, where
${\cal{V}}_{1}(y) $ is the potential. Therefore, the central problem
of the present analysis is then to find, if they exist, appropriate
self adjoint extensions of ${\cal{H}}$, where by appropriate we mean
that the resulting solutions are acceptable as perturbations, and
that any acceptable perturbation can, in principle, be expanded
using the solutions, which, in turn, provides a solution of the
evolution equations. We will elaborate further on this point in what
follows. We recall at this point that on account of (\ref{LTeq3}),
$p_1$ has the range $-2/3 \leq p_1 \leq 4/3$, and, therefore,
considering (\ref{peq30}), we see that $y$ is a monotonic function
of $x$, such that with an appropriate choice of an integration
constant, corresponding to the range $0 < x < \infty$, we have $0 <
y < y_0$, where $y_0$ is given by
\begin{equation}
\label{peq30a}
 y_0=\int_0^{\infty} \frac{2}{\sqrt{3} \,x^{p_1/2+1/3} \left(1+
 x^2\right)^{2/3-p_1/4}}dx
\end{equation}
and, therefore, is finite.

\subsection{Some properties of $ {\cal{V}}_{1}(y)$.}

As already remarked we could not find closed form expressions for
${\cal{K}}(x)$ for general values of $\kappa$. We notice, however,
that to analyze the possible behaviour of the solutions of
(\ref{peq34}) at the boundaries ($y =0$, and $y=y_0$), all that is
required are adequate expansions of $y(x)$, and of ${\cal{K}}(x)$.
Let us consider first the boundary $y=0$. In accordance with
(\ref{peq30}), (and a similar analysis carried out for Case I), near
$x=0$ we have,
\begin{equation}\label{V101}
    x \simeq \left(\frac{\sqrt{3}(4 -3 p_1)
    y}{12}\right)^{6/(4-3p_1)}
\end{equation}

A simple computation shows that near $x=0$, to leading order we
have,
\begin{equation}\label{V103}
    {\cal{V}}_{1} \simeq - \frac{(4 - 3 p_1)^2}{192} x^{p_1-4/3}
\end{equation}
and, therefore, near $y=0$, again to leading order, we have,
\begin{equation}\label{V105}
    {\cal{V}}_{1}(y) \simeq - \frac{1}{4 y^2}
\end{equation}

Thus, near $y=0$, the behaviour of the potential ${\cal{V}}_{1}(y)$
is independent of $\kappa$ and $k$. The general solution $W_4(y)$ of
(\ref{peq34}) for a potential of the form (\ref{V105}) takes the
form,
\begin{equation}\label{V107}
   W_4(y) \simeq C_1 \sqrt{y} + C_2 \ln(y)\sqrt{y}
\end{equation}
where $C_1$ and $C_2$ are arbitrary constants. It is well known that
all that is required in this case to obtain a self adjoint extension
(at least as regards the boundary $y=0$), is to keep to ratio
$C_1/C_2$ {\em at some fixed value} for all the solutions of
(\ref{peq34}). The situation is similar to that found and discussed
in our previous analysis of the perturbations of the Levi - Civita
space times \cite{glei1}. For similar reasons as those considered
there we shall impose $C_2=0$, as our boundary condition. We refer
the reader to \cite{glei1} for further details and references. We
only remark that setting $C_4 =0$ is the most restrictive condition as regards
possible eigenvalues $\Omega^2 < 0$, as can be seen by considering
the situation for a pure $-1/(4 y^2)$ potential.

The other important limit is for $ y \to y_0$, ($x \to  \infty$).
Using again the explicit form of $ {\cal{V}}_{1}$ as a function of
$x$, we find, to leading order for $x \to  \infty$,
\begin{equation}\label{V109}
    {\cal{V}}_{1} \simeq
    \frac{3(1+\kappa+\kappa^2)^2 k^6 + 2 \kappa^2 (\kappa+2)^2 \Lambda^3}{3(1+\kappa+\kappa^2)^2 k^4 \Lambda}
    + {\cal{O}}\left(x^{-2/3}\right)
\end{equation}
or, in terms of $y$,
\begin{equation}\label{V111}
    {\cal{V}}_{1}(y) \simeq
    \frac{3(1+\kappa+\kappa^2)^2 k^6 + 2 \kappa^2 (\kappa+2)^2 \Lambda^3}{3(1+\kappa+\kappa^2)^2 k^4 \Lambda}
    + {\cal{O}}\left(y_0-y\right)
\end{equation}

Therefore, the potential $ {\cal{V}}_{1}(y)$ is finite and regular
at $y=y_0$. The general solution $W_4(y)$ is then, to leading order,
of the form,
\begin{equation}\label{V113}
    W_4(y) \simeq C_1 + C_2 (y_0-y)
\end{equation}
where $C_1$ and $C_2$ are arbitrary constants. This corresponds
also, as for $y=0$, to the {\em circle limit} case for self
adjointness. However, in this case we do not have a regularity
criterion that we can apply, and, in principle, all fixed ratios
$C_2/C_1$ are acceptable, introducing an essential ambiguity in the
evolution of the perturbations. As already discussed, this can be traced to the fact that
the boundary $x \to \infty$ is time like, as discussed in
\cite{wald}. Nevertheless, if we assume that a criterion for
choosing a particular value of the ratio $C_2/C_1$ can be
established, and that that choice leads to a well defined evolution
of the perturbations, the question still remains as to which, if
any, of the possible choices leads to a stable evolution. For this
reason, in what follows we will leave the choice of the value of
$C_2/C_1$ open, and proceed to the general analysis of the solutions
of the perturbation equations. We notice, on this regards, that
(\ref{peq30}), and (\ref{peq34b}) provide a parametric (in terms of
$x$) expression for $ {\cal{V}}_{1}(y)$. We may use this fact to
gain some insight on the nature of the solutions of (\ref{peq34}),
by plotting $ {\cal{V}}_{1}(y(x))$ as a function of $y(x)$,
evaluated numerically, using,
\begin{equation}
\label{peq30f} y(x)= \int_0^{x}\frac{2}{\sqrt{3}\;x_1^{p_1/2+1/3} \left(1+
x_1^2\right)^{2/3-p_1/4}}dx_1
\end{equation}

Some examples are given in Figure 2 for several values of $\kappa$
and $k=1$. Recall that the range of $y$ depends on $\kappa$, and
that in all cases  we have $ {\cal{V}}_{1}(y(x)) \sim -1/(4 y^2) $
sufficiently close to $y=0$. The presence of a ``dip'' region, away
from $y=0$, where the potential has a negative minimum is noticeable
in all cases. We must remark here that, just as in Case I, it should
be clear that, for a given boundary condition for $y=0$, the most
restrictive condition as regards the possibility of eigenvalues
$\Omega^2 <0$, is obtained setting $C_1=0$, that is, imposing that
the solutions vanish for $y=y_0$.

\begin{figure}
\centerline{\includegraphics[height=12cm,angle=-90]{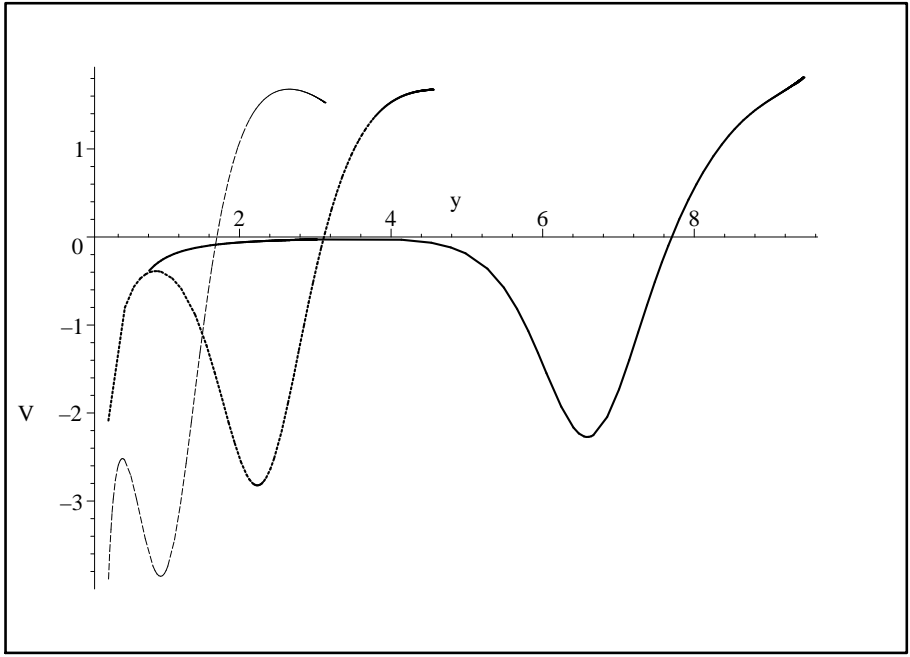}}
\caption{${\cal{V}}_{1}(y(x))$ as a function of $y$ for several
values of $\kappa$ and $k=1$. The dashed curve (leftmost)
corresponds to $\kappa=0.5$. The dotted curve (center) to
$\kappa=1$, and the solid curve (rightmost) to $\kappa=2$. Notice
the presence of a ``dip'' region in all cases, where the potential
has a minimum away from $y=0$. All the curves extend to $y=0$, and
diverge as $-1/(4y^2)$ sufficiently close to $y=0$, although this is
not shown in the plots for simplicity.}
\end{figure}

\section{The spectrum of  ${\cal{H}}$.}

Once we make appropriate choices of the boundary conditions at $y=0$
and $y=y_0$, the operator ${\cal{H}}$ in (\ref{peq34}) becomes self
adjoint and we are assured of the existence of a complete spectrum
of eigenvalues and eigenfunctions that can be used to generate the
evolution of arbitrary initial data for the perturbations. The
details are entirely similar to those analyzed in \cite{glei1}, and
will not be repeated here. The important point is that the resulting
evolution will be stable only if there are no eigenvalues with
$\Omega^2 < 0$, and, therefore, we will concentrate on finding, if
they exist, those eigenvalues and eigenfunctions. If we try to find
those solutions, either exact or numerically, directly from
(\ref{peq34}), we are faced with the difficulty that $
{\cal{V}}_{1}$ is known only implicitly as function of $y$. For this
reason, and considering that there is a one-to-one correspondence
between the solutions of (\ref{peq34}), and those of (\ref{peq28a}),
we shall analyze the solutions of the latter instead of those of
(\ref{peq34}). As a first step, we need to translate the boundary
conditions for (\ref{peq34}) into those corresponding to
(\ref{peq28a}). To achieve this we would need to solve
(\ref{peq28d}) to obtain ${\cal{K}}$, but, fortunately, we only need
its behaviour near the boundaries.

Using (\ref{peq28d}), and expanding $y(x)$ and $q_3$ to leading
order near $x=0$, after some computations we find,
\begin{equation}\label{spec01}
    {\cal{K}}(x) \simeq x^{\frac{2\kappa+1}{2(1+\kappa+\kappa^2)}}
\end{equation}
We also have $y \propto x^{\frac{1}{1+\kappa+\kappa^2}}$, so that we
finally find that near $x=0$ we should have,
\begin{equation}\label{spec03}
   W_3(x) \simeq
   x^{\frac{\kappa+1}{1+\kappa+\kappa^2}}\left(C_1+C_2
   \ln(x)\right)
\end{equation}
where $C_1$ and $C_2$ are arbitrary constants. In accordance with
our previous discussion, we must impose the condition $C_2=0$.

For the boundary at $x \to \infty$ we have,
\begin{equation}\label{spec05}
    {\cal{K}}(x) \simeq x^{-5/3}+\frac{(p_2-p_1)\Lambda}{2k^2}
    x^{-7/3}
\end{equation}

Since $x \to \infty$ we have $y \simeq y_0 -\sqrt{3} x^{-2/3}$, we find
that in that limit we should have,
\begin{equation}\label{spec07}
   W_3(x) \simeq C_1 x^{-5/3} + C_2 x^{-7/3}
\end{equation}
where $C_1$ and $C_2$ are arbitrary constants. Notice that the case
$C_1 \neq 0$, $C_2=0$ corresponds to $W_4(y)$ approaching a non zero
constant, with $d W_4/dy =0$, while for $C_1=0$, $C_2 \neq 0$, we
have $W_4(y)=0$, with $d W_4/dy \neq 0$ and finite.

There are different procedures that one can use to analyze
(\ref{peq28a}) numerically. We notice that, for a general procedure,
near $x=0$ the structure of the coefficients as functions of $x$ is
rather complicated because of the presence of exponents that are
rational functions of $\kappa$. On other hand, if we consider, for
instance, the coefficients of the unperturbed metric, for $x \to
\infty$, they all admit expansions of the form,
\begin{equation}
\label{spec09}
  x^{p_i+2/3}\left(1+x^2\right)^{1/3-p_i/2} \sim x^{4/3} \sum_{j=0} a_j
  x^{-2 j}
\end{equation}
so, that, eventually, all relevant quantities in the problem can be
expanded in (integer) powers $x^{1/3}$. In particular, it is not
very difficult, although too lengthy to be displayed here, to show
that all the coefficients in (\ref{peq28a}) admit expansions similar
to (\ref{spec09}) for sufficiently large $x$. On this account, one
can show that, for sufficiently large $x$, the general solution of
(\ref{peq28a}) admits an expansion of the form,
\begin{eqnarray}\label{spec11}
   W_3(x) &\simeq & C_1\left( x^{-5/3}+  \sum_{j=4} a_j
  x^{-(2 j+1)/3}\right) \nonumber \\
   & & + C_2 \left(x^{-7/3}+  \sum_{j=4} b_j
  x^{-(2 j+1)/3}\right)
\end{eqnarray}
where $C_1$ and $C_2$ are arbitrary constants, and the coefficients
$a_j$ and $b_j$ are functions of $\kappa$, $\Omega$, $\Lambda$, and
$k$ that can be readily computed by replacing in (\ref{peq28a}) and
expanding in powers of $x$ for $x \to \infty$. Notice that regarding
the boundary conditions for $y = y_0$, if we define $\alpha$ by,
\begin{equation}\label{spec13}
    \alpha =  \left[ W_4\left(\dfrac{d W_4}{dy}\right)^{-1}\right]_{y = y_0} ,
\end{equation}
then, we have,
\begin{equation}\label{spec15}
    \alpha = -\frac{\sqrt{3} C_1}{C_2} ,
\end{equation}
and, therefore, the ratio of the constants $C_1/C_2$ controls the
boundary value imposed at $y=y_0$.

Considering the general form of the potential, and the fact that the
range of $y$ is finite, we can see that the most restrictive
boundary condition for the existence of negative eigenvalues is
$W_4(y_0)=0$, (with $dW_4/dy|_{y=y_0} \neq 0$). This condition
corresponds to $C_1=0$, $C_2 \neq 0$.

We remark once again that we are only interested in finding whether,
after imposing boundary condition that imply that ${\cal{H}}$ in
(\ref{peq34}) is self adjoint, its spectrum contains negative
eigenvalues. This amounts to finding solutions of (\ref{peq28a}) for
negative $\Omega^2$, after imposing the corresponding boundary
conditions. Notice that, since the range of $y$ is finite, the spectrum is fully discrete,
but, because of the complicated form of the coefficients of
(\ref{peq28a}) this search can only be carried out numerically.
There are different possibilities here. We have chosen a
``shooting'' method, where, using the expansion (\ref{spec11}), we
first obtain an approximation to $W_3(x)$ and $dW_3/dx$ for a
sufficiently large value of $x$, and then integrate numerically
towards $x=0$. In more detail, after fixing the values of $\kappa$,
$k$, $C_1$, and $C_2$, (recall that we can take $\Lambda =1$,
without loss of generality), we give a tentative value for
$\Omega^2$, and integrate numerically towards $x=0$. Since for
arbitrary values of $\Omega^2$ the corresponding solution will
behave near $x=0$ as in (\ref{spec03}), a sufficiently accurate
integration should lead to a solution that approaches zero in all
cases. For this reason we look for solutions such that,
\begin{equation}\label{spec17}
    \lim_{x \to 0^+} \left[W_3(x) x^{-\frac{\kappa+1}{1+\kappa+\kappa^2}}  \right] = C
\end{equation}
where $C$ is some constant, since, as should be clear from
(\ref{spec03}), this corresponds to solutions where the terms in
$\ln(x)$ vanish. Some particular examples are shown Figures 3 to 6.
In all cases we have set $C_1=0$, as the most restrictive condition.
In Fig. 3 we show a plot of
$W_3(x)/x^{\frac{\kappa+1}{1+\kappa+\kappa^2}}$ for $\kappa=1$, and
$k=1$, for approximate solutions for the two lowest eigenvalues.
Numerically we find that increasing or decreasing $\Omega^2$
slightly from the indicated values causes the curve to diverge
either towards $+ \infty$ or $- \infty$ as we approach $x=0$, which
we interpret, as indicated above, as indicative of the existence of
an eigenvalue and corresponding eigenfunction for the type of
boundary conditions imposed on the problem. Mostly for completeness
we also show a plot of the corresponding solutions $W_3(x)$ in Fig.
4. We have explored numerically a range of values of the parameter
$\kappa$ an in all cases we have found that the lowest eigenvalue is
negative, although we have no proof that this is the case in general
for extremely large or small values of $\kappa$. Examples of these
lowest eigenvalues for different $\kappa$ are given in Fig. 5 and
Fig. 6. We also remark in all cases analyzed the lowest eigenvalue
is negative, but the next to lowest is positive, a situation that
resembles that found in \cite{glei1}.

\begin{figure}
\centerline{\includegraphics[height=12cm,angle=-90]{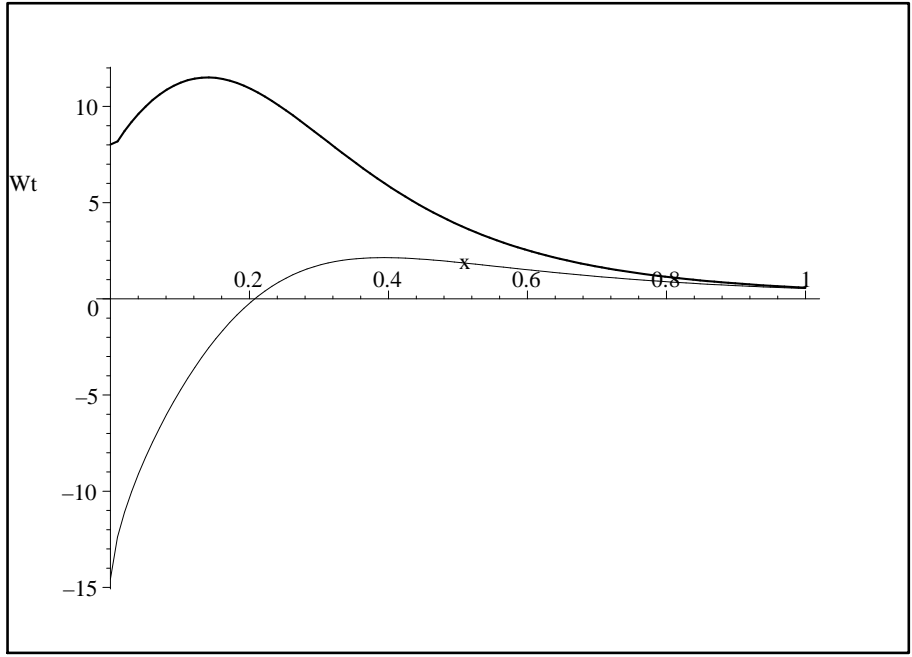}}
\caption{$W_3(x)/x^{3/2}$ as a function of $x$, for $\kappa=1$, and
$k=1$, for the two lowest eigenvalues. The thick line curve  (with
no nodes) corresponds and $\Omega^2=-1.079...$, while the thin line
curve (with one node) corresponds to $\Omega^2 =1.164...  $. The
functions are not normalized.}
\end{figure}

\begin{figure}
\centerline{\includegraphics[height=12cm,angle=-90]{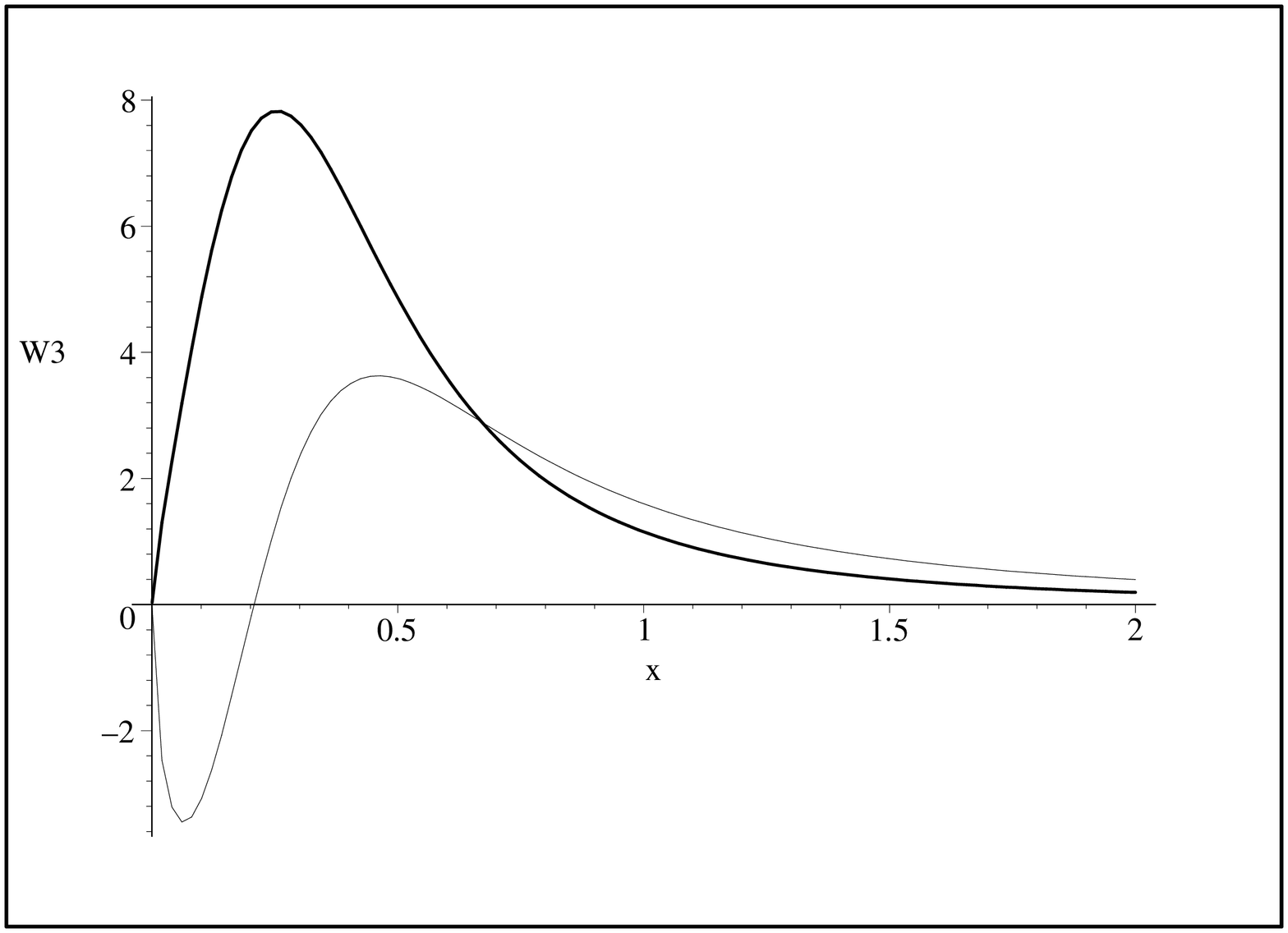}}
\caption{$W_3(x)$ as a function of $x$, for $\kappa=1$, and $k=1$,
for the two lowest eigenvalues. The thick line curve  (with no
nodes) corresponds and $\Omega^2=-1.079...$, while the thin line
curve (with one node) corresponds to $\Omega^2 = 1.164... $. The
functions are not normalized.}
\end{figure}

\begin{figure}
\centerline{\includegraphics[height=12cm,angle=-90]{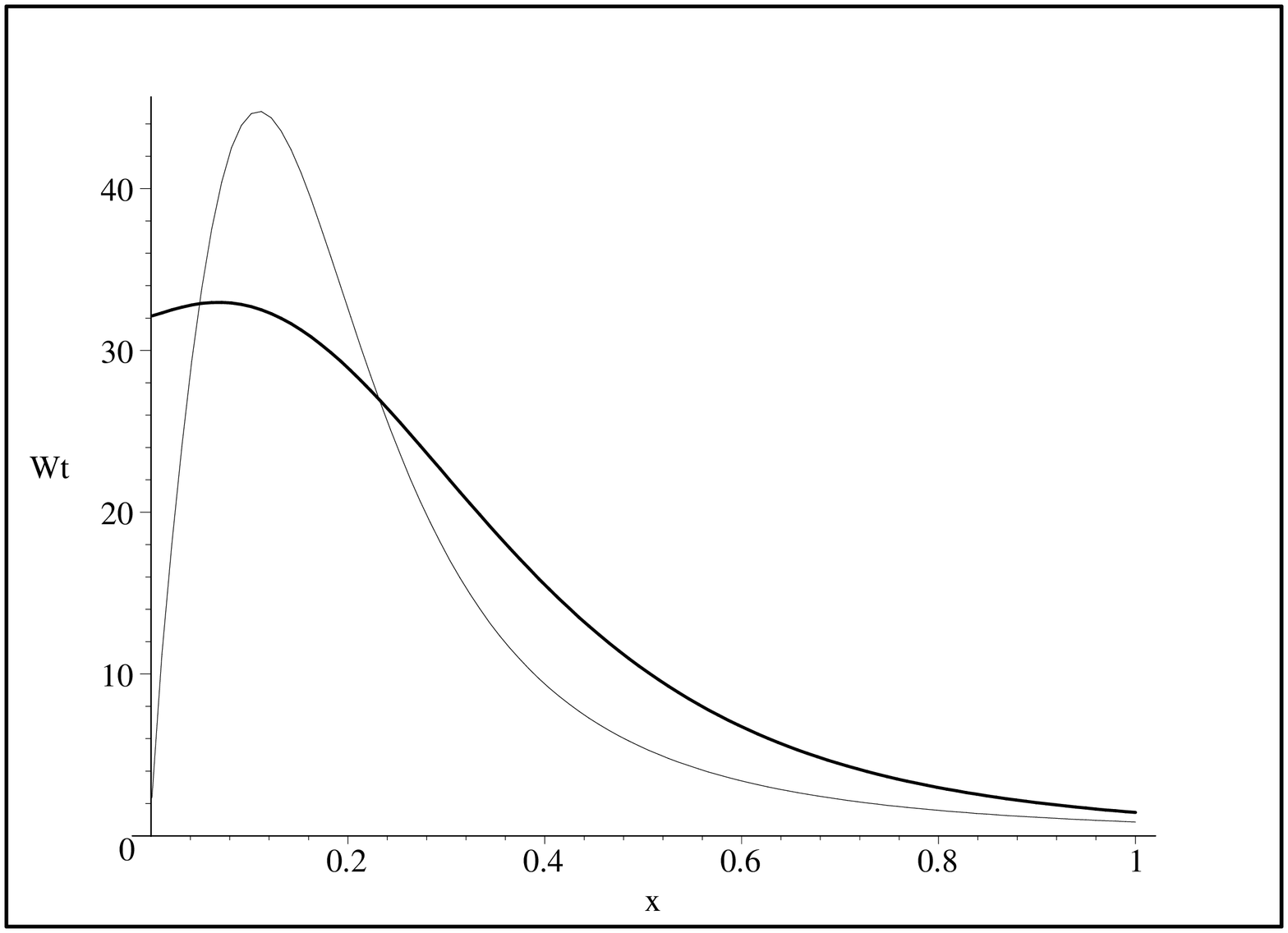}}
\caption{$W_3(x)/x^{\frac{\kappa+1}{1+\kappa+\kappa^2}}$ as a
function of $x$ for the lowest eigenvalue for $k=1$ and two
different values of $\kappa$. The thick line curve corresponds to
$\kappa=0.5$ and $\Omega^2=-1.2177...$, while the thin line curve
corresponds to $\kappa=3$, and $\Omega^2 = -0.9097...$. The
functions are not normalized.}
\end{figure}

\begin{figure}
\centerline{\includegraphics[height=12cm,angle=-90]{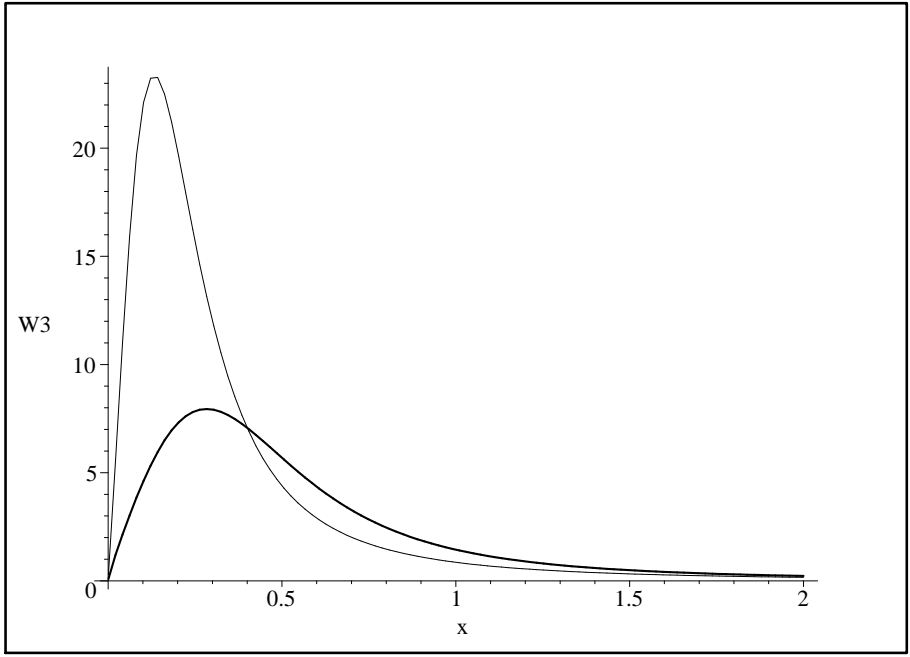}}
\caption{$W_3(x)$ as a function of $x$ for the lowest eigenvalue for
$k=1$ and two different values of $\kappa$. The thick line curve
corresponds to $\kappa=0.5$ and $\Omega^2=-1.2177...$, while the
thin line curve corresponds to $\kappa=3$, and $\Omega^2 =
-0.9097...$. The functions are not normalized.}
\end{figure}

\section{The limit $k=0$. (Radial perturbations)}

In this Section we consider the limit $k=0$. This corresponds to
purely radial perturbations that are independent of $z$, and,
therefore, preserve the cylindrical symmetry of the unperturbed
metric. One can check that the linearized Einstein equations imply
that $H_1 = H_2 = H_3 = 0$. If we consider now the linearized equations for the $F_i$, we find that in general we must have,
\begin{equation}\label{k002}
    F_7(x)=-\frac{i}{\Omega} \frac{dF_6(x)}{dx} +\frac{i(3 p_2+2+4 x^2)}{3 \Omega x (1 + x^2)} F_6(x)
\end{equation}
Without loss of generality we may write,
\begin{equation}\label{k004}
    F_6(x)=\frac{i\Omega x^{3/2+p_2}}{(1+x^2)^{p_2/2-1/3}} Z_1(x)
\end{equation}
and this implies,
\begin{equation}\label{k004}
    F_7(x)=\frac{ x^{3/2+p_2}}{(1+x^2)^{p_2/2-1/3}} \frac{dZ_1(x)}{dx}
\end{equation}
Then, in accordance with (\ref{LT06}), for $k=0$, with an appropriate choice of $Z_1(x)$, we may set $F_6(x)=F_7(x)=0$. This still leaves free the choice of $T_1(x)$ and $X_1(x)$. Considering again (\ref{LT06}), for $k=0$, we conclude that we may always choose a gauge where we also set $F_4(x)=F_5(x)=0$. Notice that $F_1(x)$ is still defined up to an additive constant. This is, of course, consistent with the resulting equations of motion. In detail we find, as independent equations,
\begin{eqnarray}
\label{k006}
  \frac{dF_1}{dx} &=& \frac{(3 p_2-4)(3 p_1-4)-32 x^4 -(24 p_1 +24p_2+32)x^2)}{6 x (1+x^2)(3p_1-4-8x^2)} F_2 \nonumber \\
   & &  +\frac{16 x^{4/3-p_1} (1+x^2)^{p_1/2+2/3} \Omega^2+ \Lambda (8 x^2+4-3 p_2)(p_1-p_2)}{x (1+x^2)(3p_1-4-8x^2)\Lambda} F_3 \; ,
\end{eqnarray}
\begin{equation}
\label{k008}
  \frac{dF_3}{dx} = \frac{(8 x^2 +4 -3 p_1)}{6 x (1+x^2)} F_2 +\frac{(p_1-p_2)}{2 x (1+x^2)} F_3 \; ,
\end{equation}
and,
\begin{equation}\label{k010}
    \frac{dF_2}{dx} = Q_1(x) F_2 +(\Omega^2 Q_2(x)+Q_3(x)) F_3
\end{equation}
where the $Q_i$ are rather complicated functions of $x$, that depend on $\kappa$, but are independent of $\Omega$.

It is now straightforward to solve (\ref{k008}) for $F_2$, replace in (\ref{k010}), and obtain a second order equation for $F_3$ of the form,
\begin{equation}\label{k012}
    \frac{d^2F_3}{dx^2} + Q_4(x) \frac{dF_3}{dx} +\Omega^2 Q_5(x)F_3+Q_6(x) F_3 =0
\end{equation}
where,
\begin{eqnarray}
\label{k014}
  Q_4(x) &=& \left[24(1+\kappa+\kappa^2)^2x^6+2(1+\kappa+\kappa^2)(4\kappa^2-2\kappa+19)x^4 \right. \nonumber \\
  & & \left.+(21-15\kappa-42\kappa^2-36\kappa^3)x^2 +9 +9 \kappa\right] \nonumber \\
  & & \times\left[x(1+x^2)(2 x^2(1+\kappa+\kappa^2)+3 +3 \kappa)(4 x^2(1+\kappa+\kappa^2)+3)\right]^{-1} \\
  Q_5 &=& \frac{4
               x^{-\frac{2 \kappa(1+\kappa)}{1+\kappa+\kappa^2}}
              (1+x^2)^{-\frac{5 +2\kappa+2\kappa^2}{3(1+\kappa+\kappa^2}}
              }{3 \Lambda} \nonumber \\
  Q_6 &=& \frac{24(\kappa+2)(2\kappa+1)\kappa}{(1+x^2)(2(1+\kappa+\kappa^2)x^2+3+3\kappa) (4(1+\kappa+\kappa^2)x^2+3)} \nonumber
\end{eqnarray}
We can use now the same techniques as in the previous Sections, and define,
\begin{equation}\label{k016}
    F_3(x)= {\cal{K}}(x) W(y(x))
\end{equation}
where $y(x)$ satisfies (\ref{peq30}), and
\begin{equation}\label{k018}
    {\cal{K}}(x) = \frac{(4(1+\kappa+\kappa^2)x^2+3) x^{-\frac{1}{2(1+\kappa+\kappa^2)}} (1+x^2)^{-\frac{1+4\kappa+4\kappa^2}{12(1+\kappa+\kappa^2)}}}{2(1+\kappa+\kappa^2)+3+3\kappa}
\end{equation}
and, one can check that this implies that $W(y)$ must satisfy the equation,
\begin{equation}\label{k020}
   -\frac{d^2 W}{dx^2} +{\cal{V}}_3(y) W(y) = \frac{\Omega^2}{\Lambda} W(y)
\end{equation}
where,
\begin{eqnarray}
\label{k022}
  {\cal{V}}_3(y) &=& \left[512\mu^4 x^8+384(2\mu+1)\mu^3 x^6-144(16\mu-15)\mu^2 x^4\right. \nonumber \\
 & & \left. -432 \mu (8\mu^2-7\mu+1)x^2-81\right]\frac{x^{-\frac{2}{\mu}} (1+x^2)^{\frac{3-4\mu}{3\mu}}}{48 \mu^2 (4\mu x^2+3)^2}
\end{eqnarray}
with $\mu =1+\kappa+\kappa^2$.

\begin{figure}
\centerline{\includegraphics[height=12cm,angle=-90]{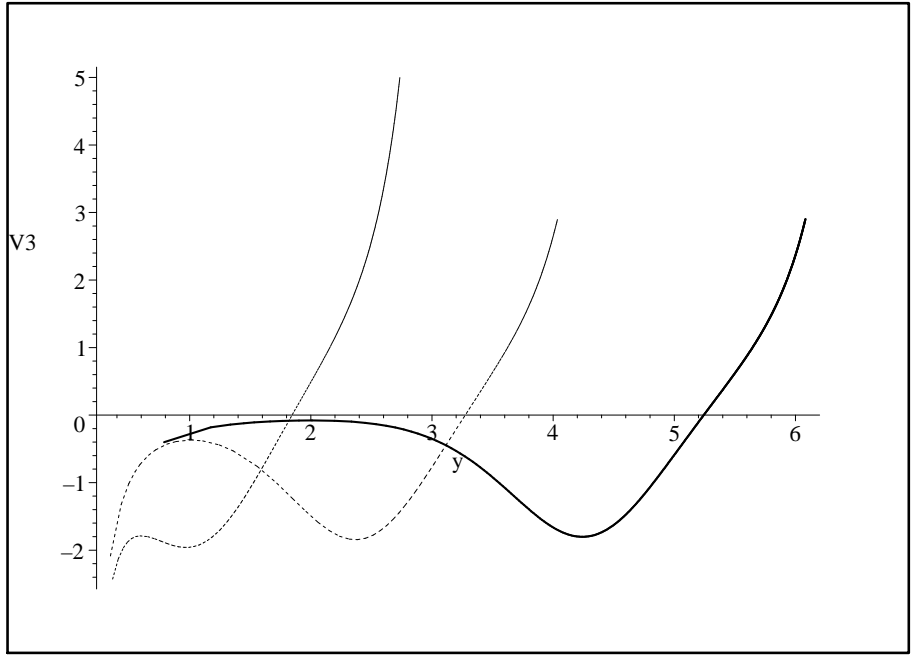}}
\caption{${\cal{V}}_3(y)$ as a function of $y$ for several values of $\kappa$. The dotted line curve (leftmost) corresponds to $\kappa=0.5$, the dashed line to $\kappa=1$ and the solid line to $\kappa=1.5$. All curves diverge as $ -1/(4 y^2)$ for $y \to 0$.}
\end{figure}

Again, as in previous Sections, we only have a rather complicated expression for ${\cal{V}}_3(y)$ as a function of $x$, and only implicitly, as a function of $y$. This still allows for a parametric representation that we can use to obtain plots of ${\cal{V}}_3(y)$ as a function of $y$, as indicated in Fig. 7. Several important points are clearly visualized in that plot. One is the divergence to $-\infty$ for $y \to 0$. Another is the divergence to $+\infty$ for $y \to y_0$, and, equally important, the ``deep'' region away from $y =0$.  We may, following similar steps as above, determine quantitatively the leading behaviour of ${\cal{V}}_3(y)$ both for $y \to 0$, (that is, for $x \to 0$), and for $y \to y_0$, (that is, for $x \to \infty$). Leaving aside details similar to those already considered, we find,
\begin{equation}\label{k024}
    {\cal{V}}_3(y) \simeq -\frac{1}{4 y^2} \;\;\;\; ; \;\;\; \mbox{for } \; y \to 0
\end{equation}
which is similar to all the previous cases. However, for $y \to y_0$, ($x \to \infty$), we find,
\begin{equation}\label{k026}
    {\cal{V}}_3(y) \simeq \frac{2}{(y_0-y)^2} \;\;\;\; ; \;\;\; \mbox{for } \; y \to y_0
\end{equation}

This immediately implies that the solutions of (\ref{k020}) behave as,
\begin{equation}\label{k028}
    W(y) \simeq \frac{C_1}{y_0-y} + C_2 (y_0-y) \;\;\;\; ; \;\;\; \mbox{for } \; y \to y_0
\end{equation}
where $C_1$, and $C_2$ are constants, and, therefore, there is a {\em unique} self adjoint extension, as far as the boundary $y=y_0$ is concerned, corresponding in this case to imposing the condition $C_1=0$ on all solutions. We complete the self adjoint specification by imposing, as before, that $W(y)\sim \sqrt{y}$, (no $\ln(y)$ terms). We can again check that these conditions translate as the following boundary conditions on $F_3(x)$:
\begin{eqnarray}
\label{k030}
  F_3(x) &\sim & C_3\left[ 1-\frac{\Omega^2 (1+\kappa+\kappa^2)^2}{3 \Lambda} x^{\frac{2}{1+\kappa+\kappa^2}}\right] \;\;\;\; ; \;\;\; \mbox{for } \; x \to 0 \nonumber \\
  F_3(x) &\sim & C_4  x^{-2} \;\;\;\; ; \;\;\; \mbox{for } \; x \to +\infty
\end{eqnarray}
where $C_3$ and $C_4$ are constants. It is again easy to obtain an asymptotic expansion for $F_3(x)$ for $x \to +\infty$, satisfying (\ref{k030}). The first terms are,
\begin{equation}\label{k032}
    F_3(x) \simeq C_4\left[\frac{1}{x^2}-\frac{3 \Omega^2}{10 \Lambda x^{10/3}}-\frac{(2\kappa^2+8\kappa+5)}{4(1+\kappa+\kappa^2) x^4} +\dots\right]
\end{equation}

We have used this expansion, together with a ``shooting'' method to obtain numerical values for $\Omega$ and solutions for $F_3(x)$, corresponding to the lowest eigenvalues and eigenfunctions in the self adjoint extension of $W(y)$. In all the cases analyzed, that correspond to a range of values of $\kappa$, we find that lowest eigenvalue corresponds to $\Omega^2 < 0$, indicating the unexpected result that the radial perturbations become {\em unstable} in the presence of a negative cosmological constant. An example is given in Fig. 8.
\begin{figure}
\centerline{\includegraphics[height=12cm,angle=-90]{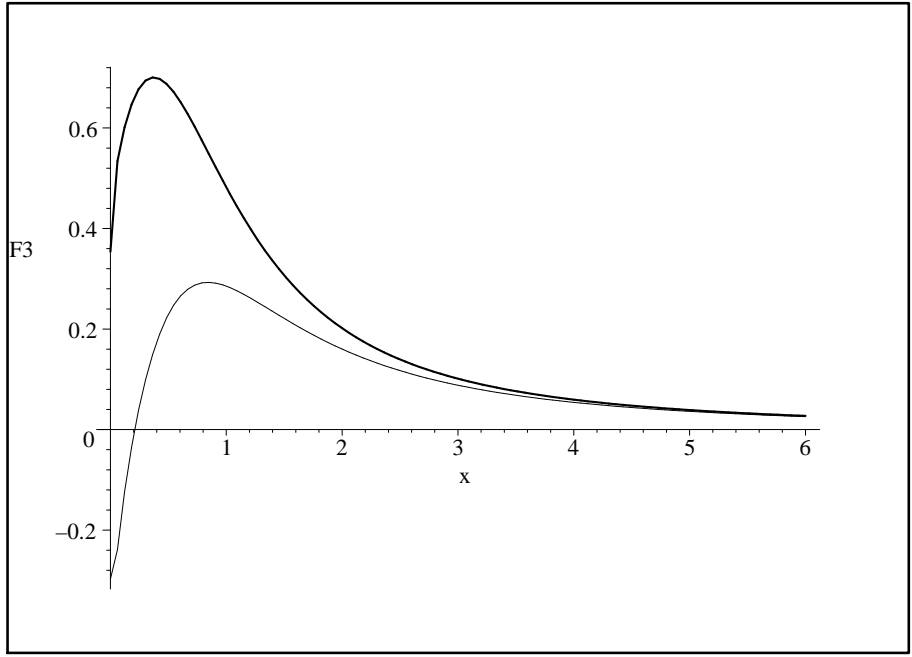}}
\caption{The solutions for $F_3(x)$, for $\kappa=1$, ($\Lambda=1$), for the two lowest eigenvalues for $\Omega$. The thick solid curve corresponds to $\Omega^2\simeq -0.57...$, and the thin solid line to $\Omega^2\simeq 1.44...$. The curves are not normalized. Notice that the curves approach a non zero value for $x \to 0$.}
\end{figure}

\section{A comment on the limit $\kappa=0$. (The Bonnor metric).}

An interesting limit of the Linet - Tian metric was analyzed by Bonnor, \cite{bonnor} where it was found that, for a negative cosmological constant, in the limit where the source vanishes, ($\kappa=0$), one obtains a static, but cylindrically symmetric anti-de-Sitter universe. We may analyze linear perturbations of that metric along the lines used in this paper. Using the gauge properties already considered we may restrict to ``diagonal'' perturbations that do not break the axial symmetry of the unperturbed metric. It is convenient in this case to write the metric (plus perturbations) in the form,
\begin{eqnarray}
\label{bon02}
  ds^2 &=& -(1+\Lambda x^2)^{2/3} \left(1 +\epsilon e^{i(\Omega t -k z)} F_1(x)\right) dt^2
  + \frac{4  \left(1 +\epsilon e^{i(\Omega t -k z)} F_2(x)\right)}{3 (1+\Lambda x^2)} dx^2 \nonumber \\
   & & (1+\Lambda x^2)^{2/3} \left(1 +\epsilon e^{i(\Omega t -k z)} F_3(x)\right) dz^2
   + \frac{x^2  \left(1 +\epsilon e^{i(\Omega t -k z)} F_4(x)\right)}{(1+\Lambda x^2)^{1/3}} d \phi^2
\end{eqnarray}

The linearized Einstein equations for this metric may be written as,
\begin{eqnarray}
\label{bon04}
  F_2(x) &=& -F_4(x) \nonumber \\
  F_1(x) &=& F_3(x)+Q \\
  F_3(x) &=& -\frac{k^2}{k^2-\Omega^2} Q+\frac{(3+2\Lambda x^2)(1+\Lambda x^2)^{2/3}}{2 x (\Omega^2-k^2)} \frac{dF_4}{dx} \nonumber \\
   & & -\frac{3 x^2 (k^2-\Omega^2)(1+\Lambda x^2)^{1/3} +2\Lambda^2 x^4 +12 \Lambda x^2+9}
       {3 x^2 (k^2-\Omega^2) (1+\Lambda x^2)^{1/3}} F_4(x)
\end{eqnarray}
where $Q$ is an arbitrary constant, and,
\begin{eqnarray}
\label{bon06}
  \frac{d^2F_4}{dx^2} &=& -\frac{1+3\Lambda x^2)}{x (1+\Lambda x^2)} \frac{dF_4}{dx}  \nonumber \\
    & &  \frac{4(3 x^2 (k^2-\Omega^2)(1+\Lambda x^2)^{1/3} +9 -2 \Lambda^2 x^4 +6\Lambda x^2)}
         {9 x^2 (1+\Lambda x^2)^2} F_4(x)
\end{eqnarray}

Therefore, every solution (\ref{bon06}) for $F_4(x)$ provides a solution for the perturbation problem. To analyze the properties of these solutions we may introduce a new function and variable defined by,
\begin{equation}\label{bon08}
    F_4(x) =K(x) W_4(y(x))
\end{equation}
where,
\begin{eqnarray}
\label{bon10}
  \frac{dy}{dx} &=& \frac{2}{\sqrt{3} (1+\Lambda x^2)^{5/6}}  \nonumber \\
   K(x) &= &  \frac{1}
         {\sqrt{x} (1+\Lambda x^2)^{1/12}}
\end{eqnarray}
and we find for $W_4$ the equation,
\begin{equation}\label{bon12}
   -\frac{d^2 W_4}{dy^2} + \frac{16 x^2 (1+\Lambda x^2)^{1/3} k^2 +40 \Lambda x^2 +45}{16 x^2 (1+\Lambda x^2)^{1/3}} W_4 (y) = \Omega^2 W_4(y)
\end{equation}
But we immediately see that the ``potential'' in (\ref{bon12}) is positive definite, and therefore, the metric is stable under this type of perturbations.

\section{A comment on the hoop conjecture.}

In interesting question regarding systems with full cylindrical symmetry, as in our case, is how the {\em hoop conjecture} manifest itself as a possible instability of the system. Here we need to remark that we are already dealing with a system that contains a naked singularity along the symmetry axis, and , therefore, it is not clear how one would assign either transverse or longitudinal radiuses to verify the conjecture. In fact, any finite section of the axis, if it could be isolated, would be in a critical situation, as its transverse radius might be thought as being zero. An important point regarding the results of our analysis is that the instabilities that we find are purely gravitational, independent of a matter contents, because they result from the presence of the ``deep'' region of the effective potential, away from the symmetry axis. Since this happens in the vacuum region, we might expect that they would remain even after the singularity is smoothed out, either by the presence of a regular cylinder of matter, or by some other way of regularizing the metric close to the symmetry axis. In this regards it is interesting to consider the behaviour of ``deep'' region as a function of $k$, since $k$ is inversely proportional to the wavelength of the perturbations in the $z$-direction. It turns out that as we increase the value of $k$ ,(shorter wavelengths), (an example is shown in Fig. 9), the ``deep'' becomes deeper and closer to the axis, so that it would be these shorter wavelengths that would be eliminated by a regularization, while the longer ones would still introduce instabilities, as, at least intuitively, expected from the hoop conjecture. In any case it would be interesting to construct explicit examples, but that is totally outside the context of the present research.

\begin{figure}
\centerline{\includegraphics[height=12cm,angle=-90]{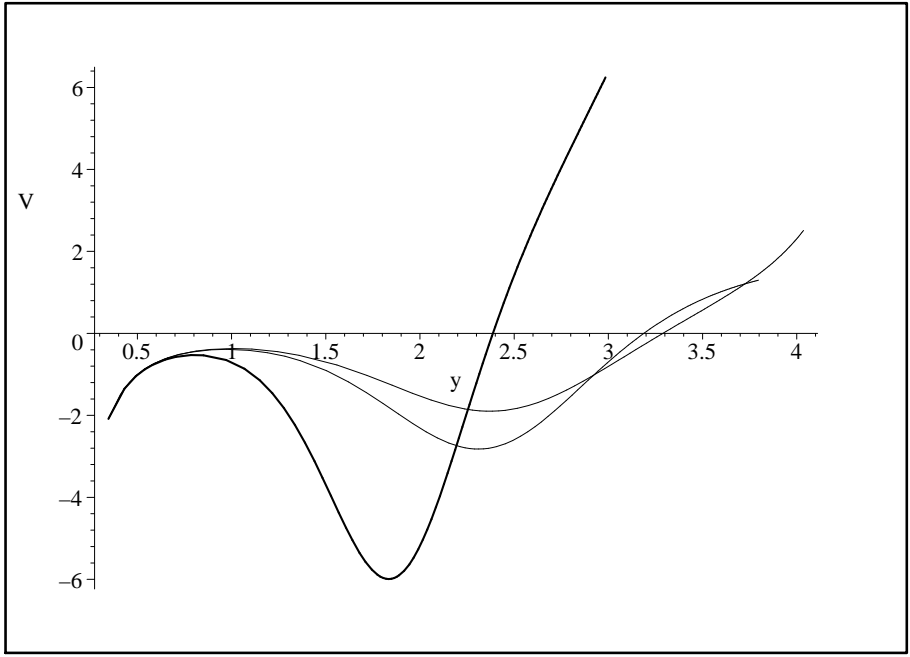}}
\caption{The potential ${\cal{V}}_1(y)$ as a function of $y$ for $\kappa=1$, for $k=0.2$, (thinner line), for $k=1$, (thin line), and for $k=4$ (thick line). The shift of the ``deep'' region towards $y=0$ with increasing $k$ is clearly seen.}
\end{figure}

\section{Final Comments}

In this paper we have analyzed the axial gravitational perturbations
of the gravitational field of an infinite line source, in the
presence of a negative cosmological constant. The analysis
generalizes a previous one where similar perturbations were studied
for the Levi - Civita metric, that corresponds to the limit of
vanishing cosmological constant. We derived the linearized equations
of motion for the perturbations and showed that they can be
separated into two sets, which we identified as Case I and Case II.
Both sets contain, in principle, a certain degree of gauge
ambiguity, which we analyzed in detail allowing us to find the
corresponding gauge invariant parts. In both cases we obtained a
``master variable'' and a corresponding ``master equation'', in the
form of a Schr\"odinger like equation. The main purpose and
usefulness of these equations is that under appropriate boundary
conditions they define a self adjoint operator, whose eigenvalues
determine the frequencies of the perturbative modes, and whose
eigenfunctions provide a complete set for the expansion and
subsequent evolution of arbitrary perturbative initial data. But,
because of the fact that for the unperturbed metric the constant $t$
hypersurfaces are not Cauchy surfaces, one needs to introduce,
somewhat arbitrarily, appropriate boundary conditions at time like
infinity in order to obtain well defined evolution equations for the
perturbations. Thus, a central problem here was establishing those
boundary conditions that fix the behaviour of the solutions at the
axis of symmetry and at infinite distance from that axis. We found
that in both cases, for the symmetry axis there is either a unique
or a physically motivated choice. On the other hand, there is,
within the context of the unperturbed metric, no unique natural or
physical choice from an infinite set of possibilities. We have
considered a set of choices, leading to a Robin type condition, that
in some way reflects that idea that ``nothing'' is added to the
initial data in its subsequent evolution. A general feature that
results is that for the assumed types of boundary conditions,
equivalent to a Robin condition on the wave equation, the resulting
evolution contains generically an unstable component. The main
result of our analysis is then that the evolution will contain
generically unstable components, and, therefore, that the space
times considered here are gravitationally unstable. Since these
space times contain a naked singularity, one would be tempted to
ascribe the instability to the presence of that singularity. We
remark, however, that here, as in the cases considered in
\cite{glei1}, \cite{Schw}, and \cite{RN}, the unstable mode is
related to the form of a ``potential'' away from the singularity,
indicating the possibility that the instability might remain even
after smoothing the (curvature) singularity by considering an
extended source. In the present case, as in \cite{glei1}, one would
have to consider a cylindrical regular source, such as an infinite
cylinder of some kind of matter. The problem in this construction is
that the resulting system is considerably more complex than the
vacuum case considered here, since, asides from  conditions that
must be imposed at the matter - vacuum boundary, we must incorporate
an equation of state for the matter, and the resulting appropriate
boundary conditions. The question, nevertheless, is interesting, but
outside the scope of the present research.

\section*{Acknowledgments}

This work was supported in part by CONICET (Argentina). I am
grateful to Gustavo Dotti for helpful comments, suggestions
and criticisms.


\begin{thebibliography}{99}
\bibitem{linet} B. Linet, J. Math. Phys. {\bf 27} 1817 (1986).
\bibitem{tian} Q. T. Tian, Phys. Rev. D {\bf 33} 3549 (1986).
\bibitem{bonnor} W. B. Bonnor, Class. Quantum Grav. {\bf 25} 225005 (2008).
\bibitem{daSilva}M. F. A. da Silva, A. Wang, F. M. Paiva and N. O. Santos, Phys. Rev. D {\bf 61}
044003 (2000)
\bibitem{griffiths} J. B. Griffiths, J. Podolsky, Phys. Rev. D {\bf 81} 064015 (2010).
\bibitem{brito} I. Brito, M. F. A. da Silva, F. C. Mena and N. O.
Santos, Gen. Relativ. Gravit. {\bf 46} 1681 (2014)
\bibitem{levi} T. Levi-Civita, Rend. Acc. Lincei {\bf 28}, 101 (1919).
\bibitem{glei1} R. J. Gleiser, Class. Quantum Grav. {\bf 32} 065003 (2015).
\bibitem{thorne} K. S. Thorne, Ph. D. Thesis, Princeton University, 1965.
\bibitem{wald} A. Ishibashi and R. M. Wald, Class. Quantum Grav. {\bf 21} 2981 (2006)
\bibitem{Schw} R. J. Gleiser and  G. Dotti,  Class. Quantum Grav. {\bf 23} (2006) 5063
\bibitem{RN} G. Dotti, R. J. Gleiser,  J. Pullin, I. F. Ranea-Sandoval and H. Vucetich, Int. J. of Mod. Phys. A, {\bf 24} (2009) 1578.
\end{thebibliography}
 \end{document}